\documentclass[12pt]{article}

\usepackage{graphicx}
\title{Bottomonium dipion transitions
}
\author{  Yu.A.Simonov and A.I.Veselov\\State Research
Center\\Institute of Theoretical and Experimental Physics, \\
Moscow, 117218 Russia}
 \date{}
\newcommand{\beq}{\begin{eqnarray}}
 \newcommand{\eeq}{\end{eqnarray}}
\newcommand{\be}{\begin{equation}}
 \newcommand{\ee}{\end{equation}}

\def\fun#1#2{\lower3.6pt\vbox{\baselineskip0pt\lineskip.9pt
\ialign{$\mathsurround=0pt#1\hfil ##\hfil$\crcr#2\crcr\sim\crcr}}}

\newcommand{{\SD}}{\rm SD}
\newcommand{{\Lc}}{\mathcal{L}}
\newcommand{{\Mc}}{\mathcal{M}}

\newcommand{\vef}{\mbox{\boldmath${\rm f}$}}

\newcommand{\veP}{\mbox{\boldmath${\rm P}$}}
\newcommand{\vep}{\mbox{\boldmath${\rm p}$}}
\newcommand{\veq}{\mbox{\boldmath${\rm q}$}}

\newcommand{\veK}{\mbox{\boldmath${\rm K}$}}

\newcommand{\vek}{\mbox{\boldmath${\rm k}$}}

\newcommand{\vepi}{\mbox{\boldmath${\rm \pi}$}}

\newcommand{\lan}{\langle}
\newcommand{\ran}{\rangle}
\begin{document}
\maketitle
\begin{abstract}
Dipion transitions of the subthreshold bottomonium levels
$\Upsilon (nS)\to \Upsilon (n'S) \pi\pi$ with  $n>n', n=2,3,4,
n'=1,2$ are studied  in the framework of the chiral decay
Lagrangian, derived earlier. The channels $B\bar B, B\bar B^*+
c.c, B^* \bar B^*$ are considered in the intermediate state and
realistic  wave functions of $\Upsilon (n S),B$ and $B^*$ are used
in the overlap matrix elements.

Imposing the Adler zero requirement on the transition matrix
element, one obtains 2d and 1d dipion spectra in reasonable
agreement with experiment.

\end{abstract}

\section{Introduction}
Dipion transitions of heavy quarkonia first discovered in
\cite{1}, were further experimentally studied in bottomonium
$\Upsilon(nS) \to \Upsilon(n'S)\pi\pi$ with  $n=2,3,~~ n'=1,2$ by
CLEO \cite{2,3,4,5}, and for $n=4, n'=1,2$ by BaBar \cite{6} and
Belle \cite{7}.

Recently a detailed analysis of dipion transitions between states
with $n=1,2,3$ was done by CLEO Collaboration \cite{8}.

On  the theoretical side the first attempt of explanation of
dipion  spectra was done in  \cite{9,10,11,12,13} using multipole
gluon field expansion and PCAC, for a recent development  of this
model see \cite{14}.

However, in this approach the natural explanation can be given
only to the $(n,n')= (2,1)$ dipion  spectrum, while other types of
spectra with a double peak need additional assumptions, such as
the role of final state interaction and $\sigma$ resonance
\cite{15,16,17,18}, exotic $\Upsilon \pi$ resonances
\cite{15,19,20,21}, coupled channel effects \cite{22,23},
relativistic corrections \cite{24}, $S-D$ mixing \cite{25}. The
role of constant term was studied in \cite{26}, for a recent
development see \cite{27,28}.

In the present paper we are using the formalism of field
correlators \cite{29} and chiral decay Lagrangian \cite{30}
developed for the dipion transitions in \cite{31,32}. In this
formalism the dipion transition proceeds via $B\bar B, B\bar B^*$
etc. intermediate states and the total amplitude consists of two
terms: $\mathcal{M}=a-b$, where $a$ refers to  the  subsequent one
pion emission at each vertex of the type $\Upsilon \to B\bar B$,
while $b$ refers to the sequence of  zero-pion and two-pion
vertices, see Fig.1. The crucial for the calculation is the
knowledge of the realistic wave functions of all participants. In
\cite{31,32} the simplest SHO form was used, fitted to the
realistic  r.m.s. of  a given state. In the present paper we are
using the realistic wave functions of $\Upsilon (nS), B$ and $B^*$
mesons, calculated in \cite{33} and being in good  agreement with
spectra and decay constants. To simplify calculations of the
overlap matrix elements we are expanding realistic wave functions
in series of SHO functions and check accuracy of expansion.


\vspace*{-4cm}
\unitlength 0.75mm 
 \linethickness{0.4pt}
\ifx\plotpoint\undefined\newsavebox{\plotpoint}\fi 
\begin{picture}(224.5,138.25)(0,0)
\put(51,51.25){\oval(56,24.5)[]} \put(134,51.75){\oval(50,25)[]}
\put(51.75,52.25){\oval(28.5,7)[]}
\put(221.25,134.5){\rule{3.25\unitlength}{3.75\unitlength}}
\put(135,52.25){\oval(27.5,7)[]} \put(37.75,52.75){\circle*{2.5}}
\put(65.5,52.5){\circle*{2.5}} \put(121.5,52.5){\circle*{2.5}}
\put(148.25,53.25){\circle*{.5}} \put(148.5,52.75){\circle*{2.55}}
\put(37.25,53.5){\circle*{1.12}} \put(38.5,52.75){\circle*{2.24}}
\put(37.43,53.18){\line(1,0){.95}}
\put(39.33,53.15){\line(1,0){.95}}
\put(41.23,53.11){\line(1,0){.95}}
\put(43.13,53.08){\line(1,0){.95}}
\put(45.03,53.05){\line(1,0){.95}}
\put(46.93,53.01){\line(1,0){.95}}
\put(48.83,52.98){\line(1,0){.95}}
\put(50.73,52.95){\line(1,0){.95}}
\put(51.68,52.93){\line(1,0){.5}}
\put(52.68,52.93){\line(0,1){0}}
\multiput(52.68,52.93)(-.0313,.0313){4}{\line(0,1){.0313}}
\put(64.93,53.18){\line(1,0){.972}}
\put(66.87,53.21){\line(1,0){.972}}
\put(68.82,53.24){\line(1,0){.972}}
\put(70.76,53.26){\line(1,0){.972}}
\put(72.71,53.29){\line(1,0){.972}}
\put(74.65,53.32){\line(1,0){.972}}
\put(76.6,53.35){\line(1,0){.972}}
\put(78.54,53.37){\line(1,0){.972}}
\put(80.49,53.4){\line(1,0){.972}}
\multiput(121.18,52.93)(.051961,.033333){15}{\line(1,0){.051961}}
\multiput(122.74,53.93)(.051961,.033333){15}{\line(1,0){.051961}}
\multiput(124.3,54.93)(.051961,.033333){15}{\line(1,0){.051961}}
\multiput(125.86,55.93)(.051961,.033333){15}{\line(1,0){.051961}}
\multiput(127.42,56.93)(.051961,.033333){15}{\line(1,0){.051961}}
\multiput(128.97,57.93)(.051961,.033333){15}{\line(1,0){.051961}}
\multiput(130.53,58.93)(.051961,.033333){15}{\line(1,0){.051961}}
\multiput(132.09,59.93)(.051961,.033333){15}{\line(1,0){.051961}}
\multiput(133.65,60.93)(.051961,.033333){15}{\line(1,0){.051961}}
\multiput(120.93,53.18)(.042484,-.033497){17}{\line(1,0){.042484}}
\multiput(122.37,52.04)(.042484,-.033497){17}{\line(1,0){.042484}}
\multiput(123.82,50.9)(.042484,-.033497){17}{\line(1,0){.042484}}
\multiput(125.26,49.76)(.042484,-.033497){17}{\line(1,0){.042484}}
\multiput(126.71,48.62)(.042484,-.033497){17}{\line(1,0){.042484}}
\multiput(128.15,47.49)(.042484,-.033497){17}{\line(1,0){.042484}}
\multiput(129.6,46.35)(.042484,-.033497){17}{\line(1,0){.042484}}
\multiput(131.04,45.21)(.042484,-.033497){17}{\line(1,0){.042484}}
\multiput(132.49,44.07)(.042484,-.033497){17}{\line(1,0){.042484}}
\put(42,25.5){\makebox(0,0)[cc]{Fig. 1(a)}}
\put(122.75,25.25){\makebox(0,0)[cc]{Fig. 1 (b)}}
\end{picture}

\vspace*{-2cm}

\bigskip
\begin{center}

Fig.1. Diagrams for pion emission from internal light quark loop.\\
Subsequent one-pion emission (a) and two-pion emission (b).

\end{center}

\bigskip

Another improvement over results of \cite{31,32} is that we also
consider transitions $(n,n')=(4,1), (4,2)$ and  compare them with
experiment \cite{6,7}.

On more fundamental level we are considering not only spectra, but
also absolute values of widths both for dipion and for pionless
decays of the type $\Upsilon (nS)\to B\bar B, B\bar B^*+cc.,
B^*\bar B^*$.

The detailed analysis, made in the paper, reveals that pionic and
pionless decays  are governed by distinct vertices, the first one
is given by the chiral decay Lagrangian (CDL) \cite{30}, while the
second one by the relativistic decay Lagrangian (RDL), being  in
some sense the relativistic generalization of the $^3P_0$ model
Lagrangian.

Correspondingly we introduce two  reasonable physical scales, $
M_{br} \approx f_\pi\approx 0.1$ GeV and $ M_\omega \approx 2
\omega$, where $\omega$ is the average energy of a light quark in
the heavy-light meson ($B$ meson), $\omega\approx 0.5$ GeV.

The resulting expressions are otherwise  parameter-free and allow
to predict the $\pi\pi$ spectrum in all transitions considered. As
was already observed in \cite{31, 32}, the form of the spectrum is
defined by the only real parameter $\eta$, which was calculated in
our approach with the account of the Adler zero  requirement (this
requirement serves as a kind of renormalization condition on
amplitudes $a,b$). We shall use below the values of Adler-zero
Improved (AZI) $\eta =\eta_{AZI}$, obtained in \cite{31,32} and
predict the $\pi\pi$ spectrum as function of total $\pi\pi$ mass
$M_{\pi\pi}\equiv q$ and angle $\theta$ of $\pi^+$ with respect to
initial $\Upsilon$ direction, making also comparison to
experiment. The plan of the paper is as follows. In section 2
general equations for dipion and pionless amplitudes are given,
taken  from  \cite{31,32}, however modified as compared to
\cite{31,32} due to appearance of two mass parameters $M_{br},
M_w$ instead of one, $M_{br}$ in \cite{31}. We also make a
correction to the Eqs. ({53,55}) of \cite{31}, valid for equal
mass case, to make it suitable  for realistic case\footnote{the
authors are grateful to Yu.S. Kalashnikova for pointing this fact
to us.}. In section 3 we describe    the calculation of overlap
matrix elements and expansion of realistic wave functions. In
section 4 and we show results of calculations for matrix elements
and spectra. In section 5 our results are discussed in comparison
with experiment. Last short section
 is devoted to conclusions and perspectives.

\section{ The bottomonium decay amplitudes}

We start with the definition of Lagrangians for the light $q\bar
q$ pair creation. As was derived in \cite{30}, the CDL has the
form (cf. Eq(39)  in  \cite{31}) in Euclidean space-time \be
\mathcal{L}_{CDL} = - i \int d^4 x \bar \psi (x) M_{br} \hat U (x)
\psi (x)\label{1}\ee where $\hat U(x)$
 is the
Nambu-Goldstone (NG) matrix \be \hat U =\exp \left( i\gamma_5
\frac{\varphi_a\lambda_a}{f_\pi}\right), \varphi_a \lambda_a
=\sqrt{2} \left( \begin{array}{ccc} \frac{\eta}{\sqrt{6}}
+\frac{\pi^o}{\sqrt{2}},& \pi^+,& K^+\\
\pi^-,&\frac{\eta}{\sqrt{6}} -\frac{\pi^o}{\sqrt{2}}, &K^o\\
K^-, &\bar K^0,& -\frac{2\eta}{\sqrt{6}}
\end{array}\right),
\label{2}\ee and $f_\pi=93$ MeV. The Lagrangian (\ref{1})
describes  the $q\bar q$ pair creation with (or without) NG meson
emission, and as will be shown below, $M_{br} \approx f_\pi$.

Another type of the $q\bar q$ pair creation occurs
 in the so-called time-turning trajectories (ttt) of a light quark
 at the boundary of the Wilson loop, when confinement ensures area
 law of the loop.  In this case  one
 can introduce an effective Lagrangian -- the Relativistic decay Lagrangian (RDL) of the form
 \be
 \mathcal{L}_{RDL} = - i \int d^4 x \bar \psi (x) M_\omega \psi
 (x)\label{3}\ee

 Here $M_\omega$ will be shown to be of the order of $2\omega$ , where $\omega$ is
 the average energy of light quark in $B$ meson, $\omega\approx
 0.5$ GeV (see Appendix 1 of \cite{31} for exact values and
 discussion). Note, that $M_\omega \gg M_{br}$ and for pionless
 decays one can neglect $ \mathcal{L}_{CDL}$ as compared to $
 \mathcal{L}_{RDL}.$
  We shall use the strong decay formalism developed in \cite{31}
  and write the amplitude $w_{nm}$ for the bottomonium transition
  from the state $n$ to the state $m$ with or without additional
  $NG$ mesons, (cf. Eq. (48) of \cite{31})\be
  w_{nm} (E) =\gamma \int \frac{d^3 \vep}{(2\pi)^3} \sum_{n_2,
  n_3} \frac{J_{nn_2n_3} J^+_{mn_2n_3}}{E-
  E_{n_2n_3}(\vep)}\label{4}\ee
  where $\gamma = \frac{M^2}{N_c}$,  $M$ is $M_{br}$  or $M_\omega$ depending on pion emission and  $J_{nn_2n_3}$ is the
  overlap matrix element of the $n$ - th state of $\Upsilon$ and
  $n_2, n_3$ states of  $B\bar B$ or $B\bar B^*$ etc. For pionless
  decay one has
  \be
  J_{nn_2n_3} (\vep) = \int \bar y_{123} \frac{d^3\veq}{(2\pi)^3}
  \tilde \psi_n (c\vep + q) \tilde \psi_{n_2} (\veq) \tilde
  \psi_{n_3}(\veq).\label{5}\ee

  Here $c=\frac{\Omega}{\Omega+\omega}\approx 1 (\Omega$ is the
  energy of heavy quark in $B$ meson, $\Omega\approx 4.83$ GeV)
 and $\bar y_{123}$ is the ratio of Dirac traces,
 $\bar y_{123} = \frac{\bar Z}{\sqrt{\prod^3_{i=1} \bar Z}}$,
 where the projection operators $\bar Z_i$ are defined in
 \cite{31} and  are of the order of 1,  so that for $\Upsilon(n)\to B\bar B$
 \be \bar y_{123} \cong\bar Z_i= tr \left(\gamma_i\frac{S_Q^+}{2\Omega}
 \gamma_5 \frac{S^-_{\bar q}}{2\omega} \frac{S^+_q}{2\omega}
 \gamma_5 \frac{S^-_{\bar Q}}{2\omega}\right)\label{6}\ee
 with $S^\pm_Q = (m_Q\pm \Omega\gamma_4 \mp i p^Q_i \gamma_i ), ~~
 S_q = m_q\pm \omega \gamma_4 \mp i p^q_i \gamma_i.$

 In Appendix 1  of \cite{31}  and Appendix 1 of the present work details of calculation of $\bar Z_i$ are given.

 As a result one obtains for $\bar y_{123}$
 $$
 \bar y_{123} \approx \bar Z = \frac{im_Q}{2\Omega^2\omega} \left[ q_i
 (2\Omega+ \omega) - p_i \frac{\omega
 \Omega}{\omega+\Omega}\right]\approx$$
 \be \approx \frac{i}{\omega} \left( q_i
 -\frac{p_i\omega}{2(\omega +\Omega)}\right).\label{7}\ee

 For the one-pion emission at the decay vertex one has instead
 \be
 J^{(1)}_{n n_2n_3} (\vep, \vek) =\int \bar y^{(\pi)}_{123}
 \frac{d^3\veq}{(2\pi)^3} \tilde \Psi_n (c\vep -\frac{\vek}{2} +
 \veq) \tilde \psi_{n_2} (\veq) \tilde \psi_{n_3}
 (\veq-\vek).\label{8}\ee

 Here $\bar y_{123}^{(\pi)}$ has the same origin as $\bar
 y_{123}$, but accounts for one pion emission at the vertex
 $\Upsilon(n) \to ( B\bar B^*+ c.c.) \pi$,
 \be \bar y^{(\pi)}_{123} = i \delta_{ik}
 \frac{m^2_Q+\Omega^2}{2\Omega^2} \frac{1} {\sqrt{2
 \omega_\pi V_3}f_\pi}\approx \frac{i\delta_{ik}}{\sqrt{2\omega_\pi
 V_3}f_\pi}.\label{9}\ee

Finally, expanding the factor $\hat U$ in (\ref{2}) to the second
 order, one obtains the overlap matrix element for the double pion
 emission vertex,
 \be
  J^{(2)}_{nn_2n_3} (\vep, \vek_1, \vek_2)= \int \bar
  y_{123}^{(\pi\pi)} \frac{d^3\veq}{(2\pi)^3} \tilde \Psi_n
  (c\vep-\frac{\veK}{2} +\veq) \tilde\psi_{n_2}(\veq)
  \tilde\psi_{n_3}(\veq- \veK)\label{10}\ee
  Here $\veK$ denotes the sum of pion momenta, $\veK= \vek_1
  +\vek_2$, and $\bar y_{123}^{(\pi\pi)}$ is
  \be
  \bar y^{(\pi\pi)}_{123} =
  \frac{i\vepi_1\vepi_2}{f^2_\pi}\frac{\bar
  y_{123}}{[2\omega_\pi(k_1) V_3
  2\omega_\pi(k_2)V_3]^{1/2}}.\label{11}\ee

 Having defined all overlap matrix elements in $w_{nm} (E)$,
 Eq.(4), one can now express total amplitudes for  processes with
 or without  pion emission. The width  of $\Upsilon(nS)$ due to
 the channel $B\bar B$ is given by the equation
 \be
 \Gamma_n =\gamma_\omega\frac{p_{BB}\tilde M_{BB}}{4\pi^2} \int
 d\Omega_p |J_{nn_2n_3} (\vep)|^2, \gamma_\omega =\frac{M^2_\omega}{N_c}\label{12}\ee

    The dipion emission amplitude consists of two terms

$$ w_{nm}^{(\pi\pi)} (E) =\gamma\left\{ \sum_k\int
\frac{d^3p}{(2\pi)^3}\frac{J_{nn_2n_3}^{(1)}(\vep,\vek_1)J^{*(1)}_{mn_2n_3}
(\vep,\vek_2)}{E-E_{n_2n_3}(\vep)-E_\pi(\vek_1)}+
(1\leftrightarrow 2)\right.$$

$$-\sum_{n'_2n'_3}\int
\frac{d^3p}{(2\pi)^3}\frac{J_{nn'_2n'_3}^{(2)}(\vep,\vek_1,\vek_2)J_{mn'_2n'_3}^*
(\vep)}{E-E_{n'_2n'_3}(\vep)-E(\vek_1,\vek_2)}-$$
\be\left.-\sum_{k^{\prime\prime}}\int
\frac{d^3p}{(2\pi)^3}\frac{J_{nn^{\prime\prime}_2n^{\prime\prime}_3}(\vep)J^{(2)*}_{mn^{\prime\prime}_2n^{\prime\prime}_3}
(\vep,\vek_1,\vek_2)}{E-E_{n^{\prime\prime}_2n^{\prime\prime}_3}(\vep)}\right\}\label{13}\ee

The probability of the process $\Upsilon (n) \to \Upsilon (n')
\pi\pi$ is obtained from $w^{(\pi\pi)}_{nn'} $ by standard rules
$$dw ((n)\to (n') \pi\pi) = |w_{nn'}^{(\pi\pi)} (E) |^2
\frac{V_3d^3\vek_1}{(2\pi)^3} \frac{V_3d^3\vek_2}{(2\pi)^3}\times
$$
\be \times  2\pi \delta (E (\vek_1, \vek_2) + E_{n'}
-E_n)\label{14}\ee and the dipion decay width is \be
\Gamma_{\pi\pi}^{(nn')} = \int dw ((n)\to (n') \pi\pi) = \int d
\Phi|\mathcal{M}|^2\label{15}\ee where $d\Phi$ is the phase space
factor \be d\Phi= \frac{1}{32\pi^3 N^2_c} \frac{(M^2+ (M')^2
-q^2)(M+M')}{4M^3}\sqrt{(\Delta M)^2 -q^2} \sqrt{q^2-4m^2_\pi}
dqd\cos \theta\label{16}\ee with the notations
$$ M\equiv M(\Upsilon(n)),~~ M'\equiv M(\Upsilon (n')),~~  \Delta
M\equiv M-M',~~$$  \be q^2\equiv M^2_{\pi\pi}= (k_1+ k_2)^2 =
(\omega_1+\omega_2)^2 - \veK^2.\label{17}\ee

The amplitude $\mathcal{M}$ can be written accordingly to
(\ref{13}) as \be \mathcal{M} =\left(\frac{M_{br}}{f_\pi}\right)^2
\bar \mathcal{M}_1- \frac{M_{br} M_\omega}{f^2_\pi} \bar
\mathcal{M}_2\label{18}\ee At this point we still do not impose on
$ \mathcal{M}$  the Adler Zero requirement.

\section{Calculation of matrix elements}

For the wave function we use the solution of the  relativistic
Hamiltonian, described in \cite{33}, (for a review see the last
ref. in \cite{29}), where the only input parameters are current
quark masses, string tension and $\alpha_s$. As a result one
obtains the bottomonium spectrum with accuracy of the order of 10
MeV and a good   agreement with  experimental lepton widths, see
Table 3 in \cite{31}.  In what follows we call this function "the
realistic wave function", meaning that it is among the best
existing ones, but the point-by-point accuracy of it is not
actually known.

 In this section we describe the method of
calculation of $\bar \Mc_1,~ \bar \Mc_2$ based on the expansion of
wave function in the full set of oscillator functions. This allows
us to do integrals in the overlap matrix elements $J,J^{(1)},
J^{(2)}$ analytically, while the specifics of wave functions  is
represented by the sequence of numbers --coefficients in  the
expansion, found by the "chi squared" procedure, namely  for any
radial  excited state wave function found in  \cite{33} we write
\be \Psi (nS, r) = \sum^{k_{\max}}_{k=1} c_k^{(n)} \varphi_k
(\beta r)\label{19}\ee where $\varphi_k (\beta r)$ is given in
Appendix 2, and oscillator parameter $\beta$ as well as
coefficients  $c_k^{(n)}$ are obtained  minimizing $\chi^2$. The
quality of approximations for different $k_{\max}$ can be seen
from the Fig.2 and 3. We also compare realistic $B$ meson wave
functions computed in \cite{33} with the one-term representation
(\ref{19}) in Fig.4. (Note, that the $B^*$ wave function is the
same as for $B$ meson in the first approximation used below).
Hence, keeping $k=1$ for $B, B^*$ mesons, the simple overlap
matrix element (\ref{5}) without $\bar y_{123}$ is \be
J^{(0)}_{n,11} (\vep)
=\int\frac{d^3\veq}{(2\pi)^3}\sum^{N_{\max}}_{k=1} c_k^{(n)}
\varphi_k (\beta_1, \veq +c\vep) \varphi^2_1 (\beta_2, \veq)=
e^{-\frac{\vep^2}{\Delta}} I_{n,11} (\vep)\label{20}\ee where
$\Delta= 2 \beta^2_1+ \beta^2_2$, and  $I_{n,11}(\vep)$ is a
polynomial in powers of $\vep^2$,  given in Appendix.

In a similar way one calculates $J^{(1)}, J^{(2)},$ \be
J^{(1)}_{n,11} (\vep, \vek) = e^{-\frac{\vep^2}{\Delta}-
\frac{\vek^2}{4\beta^2_2}} I_{n,11}(\vep)  \bar
y^{(\pi)}_{123}\label{21}\ee

\be J^{(2)}_{n,11} (\vep, \vek_1, \vek_2) =
e^{-\frac{\vep^2}{\Delta}- \frac{\veK^2}{4\beta^2_2}} ~^{(1)}
I_{n,11}(\vep)  \tilde y^{(\pi\pi)}_{123} p_i.\label{22}\ee

In (\ref{22}) we take into account that $\bar y_{123}^{(\pi\pi)}$
contains $q_i$ and $p_i$ and therefore the result of integration,
over $d^3q$ leads to the modification of
$I_{n,11}(\vep)\equiv~^{(0)}I_{n,11}(\vep)$, $I_{n,11}(\vep)\to
~^{(1)}I_{n,11}(\vep)$. All these expressions are given in
Appendix 2.

Finally the  matrix element $\Mc$ in (\ref{18}) can be rewritten
as \be \Mc= \exp
\left(-\frac{\vek^2_1+\vek^2_2}{4\beta^2_2}\right) \left(\frac{
M_{br}}{f_\pi}\right)^2 \Mc_1 -\exp \left(
-\frac{\veK^2}{4\beta^2_2}\right) \frac{M_{br}
M_{\omega}}{f_\pi^2} \Mc_2\label{23}\ee and  $\Mc_1, \Mc_2$ are
given by the expressions
$$ \Mc_1=(Z_1^*)^2 [~^{(0)} \Lc^*_{nn'}(\omega_1) +~^{(0)}
\Lc^*_{nn'}(\omega_2)]+$$ \be + (Z_1^{**})^2 [~^{(0)}
\Lc^{**}_{nn'}(\omega_1) +~^{(0)}
\Lc^{**}_{nn'}(\omega_2)];\label{24}\ee

$$ \Mc_2=Z_2^2 [~^{(1)} \Lc^*_{nn'}(0) +~^{(1)}
\Lc_{nn'}(\omega_1+\omega_2)]+$$ $$ + (Z_2^{*})^2 [~^{(1)}
\Lc^{*}_{nn'}(0) +~^{(1)} \Lc^{*}_{nn'}(\omega_1+\omega_2)]+$$ \be
+ (Z_2^{**})^2 [~^{(1)} \Lc^{**}_{nn'}(0) +~^{(1)}
\Lc^{**}_{nn'}(\omega_1+\omega_2)].\label{25}\ee Here
$~^{(k)}\Lc^s_{nn'}$ is the integral as in (\ref{13}), and $s=(),
*,**$ marks three channels: $B\bar B, B\bar B^*+cc; B^*\bar B^*$
respectively; $ k=0,1$. Exact form for $\Lc$ is
$$~^{(k)}\Lc^s_{nn'} (\omega) =\int \frac{d^3\vep}{(2\pi)^3}
e^{-\frac{\vep^2}{\beta^2_0}} \frac{~^{(k)}I_{n,11} (p)
~^{(k)}I_{n',11} (p)\left(
\frac{p}{\beta_0}\right)^{2k}}{\frac{p^2}{2 m_{BB}}-(\Delta
M^{(s)}-\omega)}=$$ \be =v\int^\infty_0 \frac{t^{k+\frac12} dt
e^{-t}~^{(k)} I_{n,11} (\sqrt{t} \beta_0)~^{(k)} I_{n',11}
(\sqrt{t} \beta_0)}{t-t^{(s)} (\omega)}\label{26}\ee

Here $v, \beta_0, t^{(s)}(\omega)$ and $\Delta M^{(s)}$ are given
in Appendix 3.

Coefficients $Z_1^{(s)}$ and $Z_2^{(s)}$ define the relative
weight of channels $s=B\bar B, B\bar B^*+c.c., B^* \bar B^*$ with
account of spin and isospin, it coincides with the corresponding
coefficients found in Table IV of \cite{34}. We obtain \be
(Z^*_1)^2= (Z_1^{**})^2 \cong 1, ~~(Z_2^{**} )^2 =7 Z^2_2,~~
(Z_2^*)2=4Z^2_2,~~ Z_2^2=\frac{\beta^2_0}{6\omega^2}\label{27}\ee

Now we can compute from $~^{(1)}I_{n,11}$ the decay width of the
$\Upsilon(nS)$ into $B\bar B, B\bar B^*+ c.c., B^* \bar B^*$ etc.
From (\ref{12}), taking into account, that $J_{nn_2n_3}(\vep)=
e^{-\frac{\vep^2}{\Delta}} ~^{(1)}I_{n,n_2
n_3}(\vep)\frac{p_i}{\omega}$, and $\lan p_i p_k\ran =\frac{1}{3}
\delta_{ik} \vep^2$, one obtains \be \Gamma(\Upsilon (nS) \to
(B\bar B)^{(s)}) = \left(\frac{M_\omega}{2\omega} \right)^2
\frac{M_B^{(s)} p^3_s}{6\pi N_c} (Z^{(s)})^2
e^{-\frac{2\vep^2_s}{\Delta} }(~^{(1)} I_{n,n_2n_3}
(p_s))^2\label{28}\ee where channel index $s=()$ for $B\bar B,~
M_B^{(*)}= \frac{2M_BM^*_B}{M_B+M^*_B}$ and for $s=**$, decay
channel is $B^* \bar B^*, M_B^{(**)} = M^*_B, (Z^{()})^2=1
(Z^{(*)})=4, (Z^{(**)})^2 =7$.

 \section{The Adler-zero improvement of  matrix elements}

 As was discussed in detail in \cite{31} the general two-pion
 amplitude, consisting of one-pion vertices as in $\Mc_1$ and
 two-pion vertex in $\Mc_2$,  satisfies the Adler zero
 requirement, i.e. vanishes for $\vek_i = \omega_i =0$, if these
 amplitudes are imbedded in the general background not
 distinguishing one- and two- pion vertices. In particular, this
 implies summing up all closed channels of the type $B\bar B$. In
 \cite{31} the AZI was realized representing the amplitude $\Mc$
 in Eq. (23) in the form
 \be
 \Mc =\exp \left( -\frac{ \vek^2_1+\vek^2_2}{4\beta^2_2}\right)
 a(\omega_1 ,\omega_2) -\exp \left(-\frac{\veK^2}
 {4\beta^2_2}\right) b(\omega_1, \omega_2)\label{28a}\ee
 $$a= \left(\frac{M_{br}}{f_\pi}\right)^2 \Mc_1,~~ b= \frac{ M_{br}
 M_\omega}{f^2_\pi}\Mc_2$$
and requiring that $a(\omega_1=0, \omega_2)= b(\omega_1=0;
\omega_2), $ and the same for vanishing $k_2 =\omega_2 =0$.

We shall now apply the AZI procedure to our decays under
consideration, $n>n', n=4,3,2$.  Before doing that, we introduce
convenient new and universal variable $x$, which is in the
interval [0,1] for all transitions. \be
x=\frac{q^2-4m^2_\pi}{\mu^2},~~ \mu^2= (\Delta E)^2 -4
m^2_\pi.\label{29}\ee

In terms of variables $x$, $\cos \theta$ the dipion decay
probability can be written as \be dw_{\pi\pi} (n,n') = C_0 \mu^3
\int^1_0 \sqrt{\frac{x(1-x)}{x+\frac{4m^2_\pi}{\mu^2}}} dx
\frac{d\cos \theta}{2} |\Mc (x, \cos \theta)|^2.\label{30}\ee

Here $C_0 =\frac{1}{32\pi^3 N^2_c} = 1.12 \cdot 10^{-4}$. Writing
two exponents in (\ref{28}) in terms of $x, \cos\theta$, one has
(cf.Eq. (90) of \cite{31}) \be \vek^2_1+ \vek^2_2 \equiv \alpha
+\gamma\cos^2 \theta  = \frac{\mu^2}{2} \left( 1-\frac{\Delta
E}{M} (1-x) + \frac{x(1-x)}{x+
\frac{4m^2_\pi}{\mu^2}}\cos^2\theta\right)\label{31}\ee \be \veK^2
=\mu^2 (1-x) (1-\delta) = \mu^2 (1-x) \left(1-\frac{\Delta
E}{M}\right)\label{32}\ee

As it was found in \cite{31}, the AZI amplitude (\ref{28})
vanishes for some value $x=\eta$, depending on the channel
($n,n')$ and this value of $\eta$ was calculated in \cite{31}.
Below we shall use these values and similarly to \cite{31}
represent the amplitude $\Mc(x, \cos \theta) $ as follows \be\Mc
(x,\cos \theta) = \Mc (x=\eta, \cos \theta) + \Mc' (x=\eta, \cos
\theta)(x-\eta)+...\label{33}\ee

Here prime denotes the derivative in $x$. It is important to note,
that as explicit calculations show  (see below), both $a$ and $b$
in (\ref{28}) do not depend appreciably on $x$, and all $x$
dependence in $\Mc$ is coming from the exponential factors. Using
(\ref{31}), (\ref{32}), one arrives at the following
representation for $\Mc$,

\be \Mc (x,\cos \theta) = \Mc_1 (\eta, \cos \theta) \left( \frac{
M_{br}}{f_\pi}\right)^2 \frac{\mu^2}{4\beta^2_2} e^{-\frac{\bar
\alpha+\bar \gamma \cos^2\theta}{4\beta^2_2}} (x-\eta)
f(\eta),\label{34}\ee where $\bar \alpha=\alpha(x=\eta),~~ \bar
\gamma=\gamma(x=\eta)$ and  $f(\eta)$ appears due to derivatives
in $x$ of (\ref{31}), (\ref{32}) \be f(x) =1 + \frac{\Delta E}{2M}
-\cos^2\theta \frac{x^2 + (1-2x) \frac{4m^2_\pi}{\mu^2}}{2 (x
+\frac{4 m^2_\pi}{\mu^2})^2}.\label{35} \ee

Thus all calculations of the dipion spectra in the AZI scheme
reduces to the calculation of $\Mc_1$ given in Eq. (\ref{24}), at
some intermediate point $x=\eta$, and subsequent integrations as
in Eq. (\ref{30}).

This procedure refers to all $(n, n')$ transitions of our set,
except for the (3,2) transition. In  the letter case the value of
$\eta=\eta_{AZI}$ found in \cite{31}, is large and negative,
$\eta\cong-3$ and expansion (\ref{33}) does not make sense for $x$
in the physical interval [0,1]. Therefore one can instead
explicitly compute $\Mc_1$ and $\Mc_2$ in (\ref{23}) as given by
(\ref{24}), (\ref{25}) and insert $M_\omega=\sqrt{2}\omega\approx
0.8 $ GeV (to be checked below by pionless decays). Computed in
this way amplitudes $\Mc_1, \Mc_2, \eta$ and resulting values of
$\Gamma_{\pi\pi}^{(n,n')}= \int dw_{\pi\pi} (n,n')$ are given in
the Table 1.
\newpage

{\bf Table 1}\\
Parameters of $(n, n')$ transitions and resulting widths
$\Gamma_{\pi\pi}(n, n')$ in comparison with  $\Gamma_{\exp}$ taken
from  \cite{35}.
 \begin{center}
\vspace{3mm}

\begin{tabular}{|l|l|l|l|l|l|} \hline

&&&&&\\

$(n,n')$& 2,1&3,1&3,2&4,1&4,2\\&&&&&\\
\hline &&&&&\\ $\mu$, GeV & 0.483& 0.85& 0.174&1.083&0.479\\&&&&&\\
\hline &&&&&\\ $\mu^2$, GeV$^2$ & 0.234& 0.721& 0.03&1.172&0.229\\&&&&&\\
\hline&&&&&\\ $\Mc_1$,
GeV$^{-1}$& -1.56&0.592&-1.66&-0.452&0.688\\&&&&&\\\hline&&&&&\\
$\Mc_2$, GeV$^{-1}$&
-0.122&-0.116&-0.322&-0.0199&-0.354\\&&&&&\\\hline &&&&&\\ $\eta$
& 0
&0.56& -2.7& 0.3& 0.61\\ &&&&&\\\hline&&&&&\\
$\Gamma/\left(\frac{M_{br}}{f_\pi}\right)^4${(keV)}& 0.132&
0.195&0.311 &0.552& 0.0071\\&&&&&\\
\hline &&&&&\\ $\Gamma_{\exp}$(keV) &6&0.56&0.9&$<2$&$<6$\\&&&&&\\

\hline

\end{tabular}

\end{center}

We turn now to  the dipion spectra as functions of $x$ and $\cos
\theta$. For the AZI decay amplitudes for the processes $(n,n')=
(2,1) , (3,1), (4,1), (4,2)$  one can use the form (\ref{35}),
however for the 2d plots in $x, \cos \theta$ one should take into
account, that $\eta$ depends on $\cos \theta$. Indeed, in the
general AZI form of the dipion decays amplitude (\ref{29}) one can
see, that the first exponent on the r.h.s depends on $\cos \theta
$ (cf. Eq. (\ref{32})) hence vanishing of $\Mc$ occurs at some
$x=\tilde \eta (cos\theta)$, where \be \tilde \eta (\cos \theta)
=\eta -\gamma \cos^2 \theta = \eta-\frac12 \frac{(1-\eta)
\eta\cos^2\theta}{\eta+\frac{4m^2_\pi}{\mu^2}}.\label{37}\ee

Correspondingly, one should replace in (\ref{35}) $\eta$ by
$\tilde \eta$ given in (\ref{37}), however this amounts to a small
correction for all four transitions under investigation.

The amplitude $\Mc(x, \cos \theta)$ can be expanded in a complete
set of polynomials in the region  $0\leq x \leq 1,  -
1\leq\cos\theta\leq 1$.  One can write for this purpose a product
of orthonormal polynomials $\bar p_n(x) p_l(z)$, where $\bar
p_n(x) =\sqrt{2n+1} P_n (2x-1), ~~ p_l (z) =
\sqrt{\frac{2l+1}{2}}P_l(z)$, and $P_k(u)$ is the Legendre
polynomial, so that \be \frac{dw}{dqd\cos \theta} =\mu^2 C_0
\sqrt{x(1-x)}|\Mc|^2=  \sum^\infty_{n, l=0} a_{nl} \bar p_n (x)
p_l(\cos \theta)\label{38}\ee \be a_{nl} =\int^1_0
dx\int^{+1}_{-1} dz \frac{dw}{dqd\cos \theta} \bar p_n (x) p_l
(z).\label{39}\ee

Tables of coefficients $a_{nl}$ for our $\Mc(x, \cos \theta)$
computed according to Eq. (\ref{39}) with $\eta \to \tilde \eta$
are given in Appendix 4.

\section{Results and discussion}
We start with the  $B\bar B$ widths, which define our only input
parameters $M_{br}$ and $M_\omega$. As shown in (\ref{29}), for
the $\Upsilon (4S) \to B\bar B$ decay one can write \be\Gamma_{4S}
(B\bar B) =\left( \frac{M_\omega}{2\omega}\right)^2 0.0033
|J^{(4)}_{BB} (p) |^2 {\rm ~ GeV}\label{40}\ee where $J^{(4)}_{BB}
(p) =\exp \left(-\frac{p^2}{\Delta}\right)^{(1)} I_{411} (p),$ and
$\Delta= 2 \beta^2_1 +\beta^2_2$  and the overlap integral
$I_{411}$ is known from the wave functions of $\Upsilon (4S)$ and
$B$ meson, which are fitted by oscillator functions with $\beta_1$
and $\beta_2$ respectively. Here $p=0.33$ GeV is the $B$ meson
momentum. Approximating $\Upsilon (4S)$ wave function with 4 and
15  oscillator functions one obtains  $J^{(4)}_{BB} (0.33)= -
0.8434$ and  -3.63 respectively, which yields $\Gamma_{4S} (B\bar
B) = 2.4$ MeV and 44 MeV.

This sensitivity  of $\Gamma_{4S}$ to the wave functions is
typical for all results of total width, both with pion emission
and without. One can exploit this fact and invert the procedure to
find the $S$ wave function (e.g. position of zeros) from the
values of widths\footnote{The authors are grateful to M.V.Danilov
for this suggestion}. As it is,  we consider the 15  term
approximation for $\Upsilon(4S)$ wave function good enough (see
Fig.3 for comparison) and can define $M_\omega$  comparing
$\Gamma_{theor}$ with $\Gamma_{\exp} (\Upsilon (4S) \to B\bar
B)=(20.5\pm 2.5)$ MeV \cite{35}, which yields
$\frac{M_\omega}{2\omega} \simeq 1.46$ or $M_\omega \approx 1.72$
GeV with accuracy of $\pm 10\%$. However, the accuracy of the wave
function calculation in \cite{33} and its simulation by oscillator
functions is not well known, and to be on the safe side, we
conclude that $M_\omega\approx 2 \omega $ as an order of magnitude
estimate. We now turn to the total dipion widths
$\Gamma_{\pi\pi}(nn')$ given in Table 1. Comparing two last lines
in the Table for $\Gamma_{theor}$ and $\Gamma_{exp}$ one can see
that one gets correct order of magnitude for
$\frac{\Gamma_{theor}}{\left( \frac{M_{br}}{f_\pi}\right)^4},$ so
that the factor $\left( \frac{M_{br}}{f_\pi}\right)^4$ can be
taken in the interval $1\leq \left(
\frac{M_{br}}{f_\pi}\right)^4\leq 6$, i.e. $1\leq
\frac{M_{br}}{f_\pi}\leq 1.45$. A large discrepancy for
$\Gamma(21)$ in  Table 1 possibly is due to poor approximation of
$\Upsilon(2S)$ wave function.

In this way we are supporting our assumption, that string decay
with and without pion emission is governed by two different scale
parameters $M_{br}$ and $M_\omega$, which differ by one order of
magnitude, $M_\omega\approx 6\div 10  M_{br}$. Moreover each of
the parameters is defined by its dynamical mechanism: $M_\omega
\approx 2\omega$ due to time-turning trajectories, and $M_{br}\sim
f_\pi$ due to chiral decay Lagrangian.

Now we turn to the dipion spectra. For each transition $(nn')$ we
show in Figs. 5-15  three spectra:$ \frac{dw}{dq}, \frac{dw}{d\cos
\theta}$ and $\frac{d^2 w}{dqd\cos \theta}$, where the 1d spectra
are integrated over another variable. For comparison the existing
experimental spectra  are given in Fig.16-18,  taken from
\cite{6,7,8}  with theoretical curves from \cite{31,32} where
$\eta$ is independent of $\cos \theta$.

One can see   a general qualitative and semiquantitative agreement
between theoretical and experimental spectra. One should stress
again, that while the total dipion width is sensitive to the  form
of wavefunctions, the qualitative form of the  spectrum is defined
by the value of $\eta\approx \tilde \eta$, which accumulates the
information on overlap matrix elements  and stabilized when the
AZI  condition is imposed. The case of the (3.2) transition has a
special feature of low available phase space,  and therefore the
spectrum as function of $x$ is defined mostly by the phase space
factor $[x(1-x)]^{1/2}$, while dependence on $\cos \theta$ is also
weak due to small $\mu^2=0.03$ GeV$^2$.

Thus all features of spectra and total widths can be understood in
terms of the  formalism, presented in \cite{31,32} and in this
paper above. Note, that our more refined analysis in the present
paper has required two important changes as compared the formalism
in \cite{31,32}. Firstly, two independent decay scales $M_{br}$
and $M_\omega$ are introduced and estimated here in contrast to
the only parameter $M_{br}$ in \cite{31,32}. Secondly, the
correction was introduced in the $P$-wave vertex, viz. $\bar
y_{123}$ in Eq. (\ref{7}), which  decreases the effective value of
this vertex. However, in the AZI form the $P$-wave vertices (the
amplitude $\Mc_2$ in (\ref{23})) are derived from the Adler zero
condition, as in (\ref{28},)(\ref{34}) and the  resulting form  of
the total AZI amplitude is the same as in  \cite{31,32} (with
additional  weak $\cos \theta$ dependence in $\tilde \eta
(\cos\theta)$ neglected there).

\section{Conclusions and outlook}

We have derived the amplitudes, spectra and total widths of dipion
emission in all subthreshold processes $\Upsilon(nS) \to \Upsilon
(n'S)\pi\pi$, with $n=2,3,4$ and $n'=1,2$. We have shown that the
formalism of pair creation, based on two effective Lagrangians.
CDL and RDL with two mass parameters, $M_{br}$ and $M_\omega$
respectively, can successfully describe the experimentally found
spectra. Moreover, it was shown that two distinct values of
widths,\\ $\Gamma_{nS} (B\bar B)=O(10$ MeV) and $\Gamma_{\pi\pi}
(nn') =O(1$ keV) can be explained by much different scales of
$M_\omega\approx 2\omega=0(1 GeV)$ and $M_{br}\approx f_\pi=0
(100$ MeV). Fixing in this way $M_\omega$ and $M_{br}$ one obtains
parameter- free one- and two-dimensional spectra in good
qualitative and semiquantitative agreement with experimental data.

The application of this formalism to the  case of $\Upsilon (5S)$
states and dipion transitions to $\Upsilon (2S), \Upsilon(1S)$ is
straightforward and is now under investigation \cite{36}. The main
difference is that the $5S$ state is above the $BB,  BB^*+c.c.,
B^*B^*$ thresholds and hence the amplitudes $\Mc_1, \Mc_2$ acquire
large imaginary parts, which strongly deform the spectra and are
affected by final state $\pi\pi$ interaction.

Another direct application of formalism  to the  one-pion  or
$\eta$ transitions is clearly visible and can be a good check of
the approach, planned for the future.

The authors are grateful to many useful remarks and suggestions of
participants of seminars at ITEP (Moscow) and ITP (Heidelberg).

Support and suggestions of M.V.Danilov and S.I.Eidelman are
gratefully acknowledged. Data, advices and help from A.M.Badalian
and B.L.G.Bakker were very important for the authors. This work
was financially supported by the grants RFFI  06-02-17012 and
NSh-4961.2008.2 , and also  grant RFFI 06-02-17120.

\vspace{2cm}

{\bf Appendix 1}\\

{\bf Calculation of the coefficient $\bar y_{123}$ in Eq. (\ref{5})}\\

 \setcounter{equation}{0} \def\theequation{A1.\arabic{equation}}

General formalism for the calculation of the overlap matrix
elements and, in particular, of the factor $\bar y_{123}$ in
(\ref{5}) is given in Appendix 1 of \cite{31}. Here we present the
corrected form, where all relative momenta are properly defined.
One starts with the calculation of the trace, Eq. (\ref{6}), which
gives (cf. Eq.({52}) of \cite{31}). \be \bar Z_i =\frac{i
m_Q}{2\Omega\omega} \left( p_{qi} - p_{\bar q i} +
\frac{\omega}{2\Omega}(P_{Q_i}-P_{\bar P_{ Qi}}) +
O\left(\frac{1}{\Omega^2}\right)\right)\label{A1.1}\ee

For the $(Q \bar q))$ system one can   define total and relative
momenta as\be \veP_1 =\vep_Q + \vep_{\bar q},~~ \veq_1
=\frac{\omega_{\bar q} \vep_Q -\Omega_Q \vep_{\bar
q}}{\omega_{\bar q}+\Omega_Q}\label{A1.2}\ee and the same for the
$(\bar Q q)$ system.

\be \veP_2 =\vep_{\bar Q} + \vep_q,~~ \veq_2 =\frac{\omega_q
\vep_{\bar q} -\Omega_{\bar Q \vep_q}}{\omega_q +\Omega_{\bar
Q}}.\label{A1.3}\ee

In the total c.m. system $\veP_1 = \vep,~~ \veP_2 = -\vep$ and \be
\vep_{\bar q}= - \veq_1
+\frac{\omega}{\omega+\Omega}\vep\label{A1.4}\ee

\be \vep_{ q}= - \veq_2 -\frac{\omega}{\omega+\Omega}\vep;~~
  \vep_Q=\vep-\vep_{\bar q}; ~~ \vep_Q=\vep_{\bar Q} = -\vep-\vep_q. \label{A1.5}\ee

  Finally one obtains from (\ref{A1.1}), taking into account that
  $\veq_2 =- \veq_1\equiv -\veq$
  \be
  \bar y_{123} \approx \bar Z_i = \frac{i m_Q}{2\Omega^2 \omega} \left [
  q_i (2\Omega +\omega) - p_i \frac{\omega \Omega}{\omega
  +\Omega}\right]. \label{A1.6}\ee

  The appearance of the term $ O(q_i)$ in (\ref{A1.6}) leads to
  the change of the result of integration  over $d^3\veq$ in  Eq.
  (\ref{5}) as compared to the standard  integral $~^{(0)}I_{n,11}
  (p)$, given in (\ref{A2.10}). Separating the factor
  $\frac{p_i}{\omega}$, one arrives  at the expression
  $$
  ~^{(1)} J_{n,11} (\vep) =\int \bar y_{123} (p, q) \frac{d^3
  q}{(2\pi)^3} \psi (nS, \veq+\vep) \psi^2_{HL} (\veq)=$$
  \be
  = \frac{ip_i}{\omega} \sum_k  c^{(n)}_ka_k \left( -
  \frac{2}{\Omega+\omega}\left( \frac{\omega}{2}
  +\Omega\frac{\beta^2_2}{\Delta_n}\right) + \frac{8\beta^2_1
  \beta^2_2 \partial}{\Delta^2_n y \partial f^2}\right) \Phi
  (-(k-1), \frac32, f^2)\label{A1.7}\ee

  Here $a_k, \beta_1, \beta_2, \Delta_n, y$ are  defined in
  Appendix 2, and $c^{(n)}_k$ are coefficients of expansion of
  $\psi(nS, \veq)$ over oscillator functions, see Appendix 2 for
  more details.

\vspace{2cm}

{\bf Appendix 2}\\

 \setcounter{equation}{0} \def\theequation{A2.\arabic{equation}}

The SHO basis functions $\{ \varphi_k (\beta r)\}$ can be written
as \be \varphi_k (\beta, r) = c_k \frac{H_{2k-1} (\beta r)}{\beta
r} e^{-\frac{(\beta r)^2}{2}}\label{A2.1}\ee

With the normalization condition $\int \varphi_k (z) \varphi_{k'}
(z) d^3 z = \delta_{kk'}$ and coefficients \be
c_k=\frac{1}{(2^{2k} \pi^{3/2} (2k-1)!)^{1/2}} \label{A2.3}\ee One
can write Hermite polynomials in (\ref{A2.1}) as

\be \frac{H_{2n-1}(x)}{x} = (-1)^{n-1} 2 \frac{(2n-1)!}{(n-1)!}
\Phi (-(n-1), \frac32, x^2)\label{A2.4}\ee

where \be \Phi(\alpha,\gamma, x^2)  = 1+ \frac{\alpha}{\gamma}
\frac{x^2}{1!} + \frac{\alpha(\alpha+1)}{\gamma(\gamma+1)}
\frac{(x^2)^2}{2!}+... \label{A2.5}\ee

Any radial excited state $nS, n=1,2,3,...$ can be expanded as \be
\psi(nS; \beta,r) = \sum^\infty_{k=1} c_k^{(n)} \varphi_k (\beta
r); \label{A2.6}\ee the overlap matrix element can be written as
\be J_{n;1,1} (p) \equiv \int \frac{d^3 q}{(2\pi)^3} \psi (nS;
\veq+ \vep) \psi^2_{HL} (1S; \veq)\label{A2.7}\ee

where the heavy light meson wave function is \be \psi_{HL} (n'S;
\beta_2, r) =\sum_k \bar c_k^{(n')} \varphi_k
(\beta_2,r)\label{A2.8}\ee $$ J_{n;1,1} (\vep) =\sum_{k(k_1, k_2)}
c_k^{(n)} \bar c^{(1)}_{k_1} \bar c^{(1)}_{k_2} \int
\frac{d^3q}{(2\pi)^3} \varphi_k(\beta_1, \bar q+ \bar p)\times $$
\be \times \varphi_{k_1}(\beta_2, \veq)  \varphi_{k_2} (\beta_2,
\veq)\label{A2.9}\ee
$$
\int\frac{d^3 q}{(\pi)^3} \varphi_k (\beta_1, \bar q +\vep)
\varphi_1 (\beta_2, \veq)\varphi_1 (\beta_2, \veq)= $$
$$=e^{-\frac{\vep^2}{\Delta}} ~~ ^{(0)}I_{n,1,1}(\vep) =
e^{-\frac{\vep^2}{\Delta} } \tilde c_n(-1)^{n-1} 2
\frac{(2n-1)!}{(n-1)!} \Phi(-(n-1), \frac32, f^2)\times$$ \be
\times \frac{y^{n-1}}{(2\sqrt{\pi})^3}
\left(\frac{2\beta^2_1\beta^2_2}{\Delta_n}\right)^{3/2}\label{A2.10}\ee
where

$$\Delta_n=2\beta^2_1+\beta^2_2, ~~
y=\frac{2\beta^2_1-\beta^2_2}{2\beta^2_1+\beta^2_2},~~  \vef =
\frac{2\vep\beta_1}{\Delta_n \sqrt{y}};$$

$$ \tilde C_n =\left(\frac{2\pi}{\beta_1}\right)^{3/2}
\frac{(2\sqrt{\pi}/\beta_2)^3}{(2^{2n} \pi^{3/2} (2n-1)!)^{1/2}}$$

\vspace{2cm}

{\bf Appendix 3}\\

{\bf Coefficients of the overlap integrals $~^{(k)}\Lc_{nn'}^{(s)} (\omega)$}\\
\setcounter{equation}{0} \def\theequation{A3.\arabic{equation}}

For the given intermediate state $s=(), *, **$ thresholds $E_s$
are $2M_B, M_B+M_{B^*}, 2 M_{B^*}$ respectively and one can define
$ \Delta M^{(s)} = E_s-M$, where $M$ is the mass of
$\Upsilon(nS)$. The resulting values of $\Delta M^{(s)}_{nn'}$ are
given below in the Table 2. We also assemble here some  formulas
for the quantities  appearing in Eqs. (\ref{25}), (\ref{26}).
Namely \be v\equiv \frac{2 \tilde
M^{(s)}\beta_0}{(2\pi)^2},~~\beta^2_0
=\frac{\Delta_n\Delta_{n'}}{\Delta_n+\Delta_{n'}},~~ \Delta_n = 2
\beta^2_1 + \beta^2_2, ~~ \Delta_{n'}= 2\beta'_1 +
\beta^2_2\label{A3.1}\ee Where $\beta_1, \beta'_1$ are oscillator
parameters found from fitting of realistic $(nS), (n'S)$ wave
functions with series of oscillator wave functions. Both $\beta_1$
and $\beta'_1$ depend on  number $k_{\max}$ of functions kept in
the series.
\be \tilde M^{(s)} =\frac12 M_B,~~ \frac{M_B
M_{B^*}}{M_B+M_{B^*}},~~ \frac12 M^*_B~~{\rm for}~~ s=(~), *,
**,\label{A3.2}\ee \be t^{(s)} (\omega) =\frac{2\tilde
M^{(s)}}{\beta^2_0} (\Delta M^{(s)} -\omega), s=(),
*,**.\label{A3.3}\ee

Other useful relations are \be k_1^2+k^2_2 =\alpha + \gamma\cos^2
\theta =\frac{\mu^2}{2} \left(1 -\frac{\Delta E}{M} (1-x)+
\frac{x(1-x)}{x+\frac{0.08}{\mu^2}}\cos
\theta\right)\label{3.4}\ee \be (\vek_1 +\vek_2)^2 =\mu^2 (1-x)
\left(1-\frac{\Delta E}{M}\right)\label{A3.5}\ee \be \omega_{1,2}
=\frac{\Delta E}{2} -\frac{\mu^2(1-x)}{4M} \mp \frac{\mu}{2}
\left(1-\frac{\Delta E}{2M}\right) \left(
\frac{x(1-x)}{x+\frac{0.08}{\mu^2}}\right)^{1/2}\cos
\theta\label{A3.6}\ee

{\bf Table 2}\\ Values of mass parameter  in
(\ref{23})-(\ref{26}).

\begin{center}

\begin{tabular}{|c|c|c|c|c|c|}\hline&&&&&\\
 $(n/ n')$& 21&31&32&41&42\\
\hline&&&&&\\
$\mu$(GeV)&0.483&0.85&0.174&1.083&0.479\\
\hline&&&&&\\
$\mu^2$ GeV$^2$ &0234&0.721&0.03&1.172&0.229\\
\hline&&&&&\\
$\Delta E,$ GeV& 0.56&0.895&0.332&1.12&0.556\\
\hline&&&&&\\
$\Delta M_{nn'}$ GeV&-0.54&-0.205&-0.205&0.02&0.02\\
\hline&&&&&\\
$\Delta M^*_{nn'} $ GeV&-0.58&-0.25&-0.25&-0.026&-0.026\\\hline&&&&&\\
$\Delta M^{**}_{nn'}$ GeV & -0.625&-0.295&-0.295 &-0.072&-0.072\\
&&&&&\\\hline

\end{tabular}

\end{center}

\vspace{2cm}

{\bf Appendix 4}\\

{\bf Coefficients of the 2d expansion of the probability  }
\\
\setcounter{equation}{0} \def\theequation{A4.\arabic{equation}}

As is given in (\ref{38}), (\ref{39}), the full probability
$\frac{dw}{dqd \cos \theta}$ can be expanded in products of
polynomials \be \frac{dw}{dqd \cos \theta} =\sum^{n_{\max,}
l_{\max}}_{n,l=0} a_{ln} \bar p_n (x) p_l (\cos
\theta)\label{A4.1}\ee with \be \bar p_n(x) = \sqrt{2n+1} P_n
(2x-1), p_l (z) =\sqrt{\frac{2l+1}{2}} P_l(z).\label{a4.2}\ee

The coefficients $a_{nl}$  calculated for $\frac{dw}{dqd \cos
\theta}$ in (\ref{38}), for transitions $(n, n') =(2,1), (3,1),
(4,1), (4,2)$ are given below together with $a_{nl}$ for the (3,2)
transition, calculated as in (\ref{23}). Note that $a_{ln}\equiv0$
for odd $l$.

\bigskip
{\bf Table 3}~~ The coefficients $a_{ln}$ defined as in (\ref{A4.1}) for the transition (2,1)\\
\begin{center}

\begin{tabular}{|l|c|c|c|c|c|}\hline~~~~~n&&&&&\\
  & 0 &         1 &         2 &         3 &4\\
l&&&&&\\
\hline&&&&&\\ 0&0.3890825&  0.2707011& -0.0407149& -0.0996852&
-0.0427755\\

\hline&&&&&\\
2 &-0.1300538& -0.0895548&  0.0148794 & 0.0335517&  0.0138329\\
\hline&&&&&\\
4 & 0.0109938  &0.0072973 &-0.0016296& -0.0029011& -0.0010309\\
\hline

\end{tabular}

\end{center}

\newpage

\bigskip
{\bf Table 4} ~~The same as in Table 3 for the transition (3,1)\\
\begin{center}

\begin{tabular}{|l|c|c|c|c|c|}\hline~~~~~n&&&&&\\
  & 0 &         1 &         2 &         3 &4\\
l&&&&&\\
\hline
 &&&&&\\
0& 0.2054341& -0.0268913&  0.1136786&  0.0082049& -0.0731623\\

\hline&&&&&\\
2& -0.0739902 & 0.0651164& -0.0402644& -0.0209861& 0.0284658\\

\hline&&&&&\\
2&  0.0145838& -0.0245765&  0.0033265&  0.0089023& -0.0048144\\
\hline

\end{tabular}

\end{center}

\bigskip
{\bf Table 5} ~~The same as in Table 3 for the transition (3,2)\\
\begin{center}

\begin{tabular}{|l|c|c|c|c|c|}\hline~~~~~n&&&&&\\
  & 0 &         1 &         2 &         3 &4\\
l&&&&&\\
\hline &&&&&\\ 0&4.4761992 & 0.1298668 &-1.2300728& -0.0479114&
-0.1895183\\
\hline&&&&&\\
  2& -0.0037809& -0.0002265 & 0.0005379&  0.0001151&  0.0003569\\
\hline&&&&&\\
4& -0.0248627 &-0.0007194&  0.0068405&  0.0002649&
0.0010495\\\hline

\end{tabular}

\end{center}

\bigskip
{\bf Table 6} ~~The same as in Table 3 for the transition (4,1)\\
\begin{center}

\begin{tabular}{|l|c|c|c|c|c|}\hline~~~~~n&&&&&\\
  & 0 &         1 &         2 &         3 &4\\
l&&&&&\\
\hline
&&&&&\\
0&  0.6245209  &0.5257702&  0.0810291& -0.1920402& -0.1252025\\
\hline
&&&&&\\
2& -0.0786476& -0.0707901 &-0.0530417&  0.0376144&
0.0335737\\

\hline&&&&&\\
 4& -0.0212058 &-0.0278274&  0.0141198& 0.0064219 &-0.0028386\\
\hline

\end{tabular}

\end{center}

\bigskip
{\bf Table 7} ~~The same as in Table 3 for the transition (4,2)\\
\begin{center}

\begin{tabular}{|l|c|c|c|c|c|}\hline~~~~~n&&&&&\\
  & 0 &         1 &         2 &         3 &4\\
l&&&&&\\
\hline
&&&&&\\
0&  0.0165037& -0.0083040 & 0.0080081 & 0.0029911& -0.0055958\\

\hline &&&&&\\2&-0.0063807 & 0.0055190& -0.0024492 &-0.0020172&  0.0019898\\
\hline&&&&&\\
  4&
   0.0010072 &-0.0012899 & 0.0000858 & 0.0004939& -0.0002193 \\\hline

\end{tabular}

\end{center}

\begin{center}

\includegraphics[width=7cm,keepaspectratio=true]{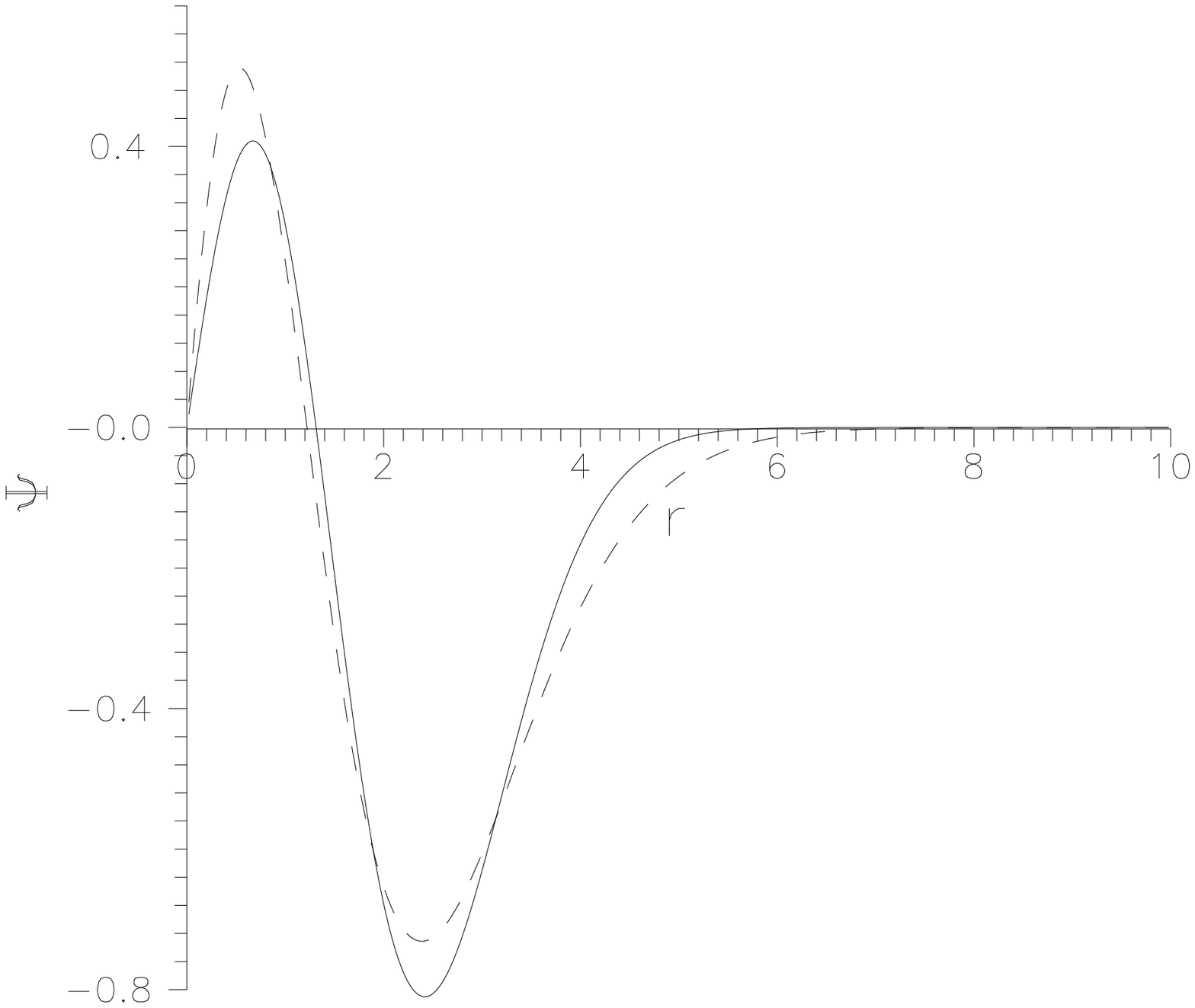}
\bigskip

 Fig.2. Realistic w.f. of $\Upsilon (2S)$  from [33] (dashed line) and
series of oscillator functions (19) with  $ k_{\max}=2$.

\end{center}

 \begin{center}

\includegraphics[width=7cm,keepaspectratio=true]{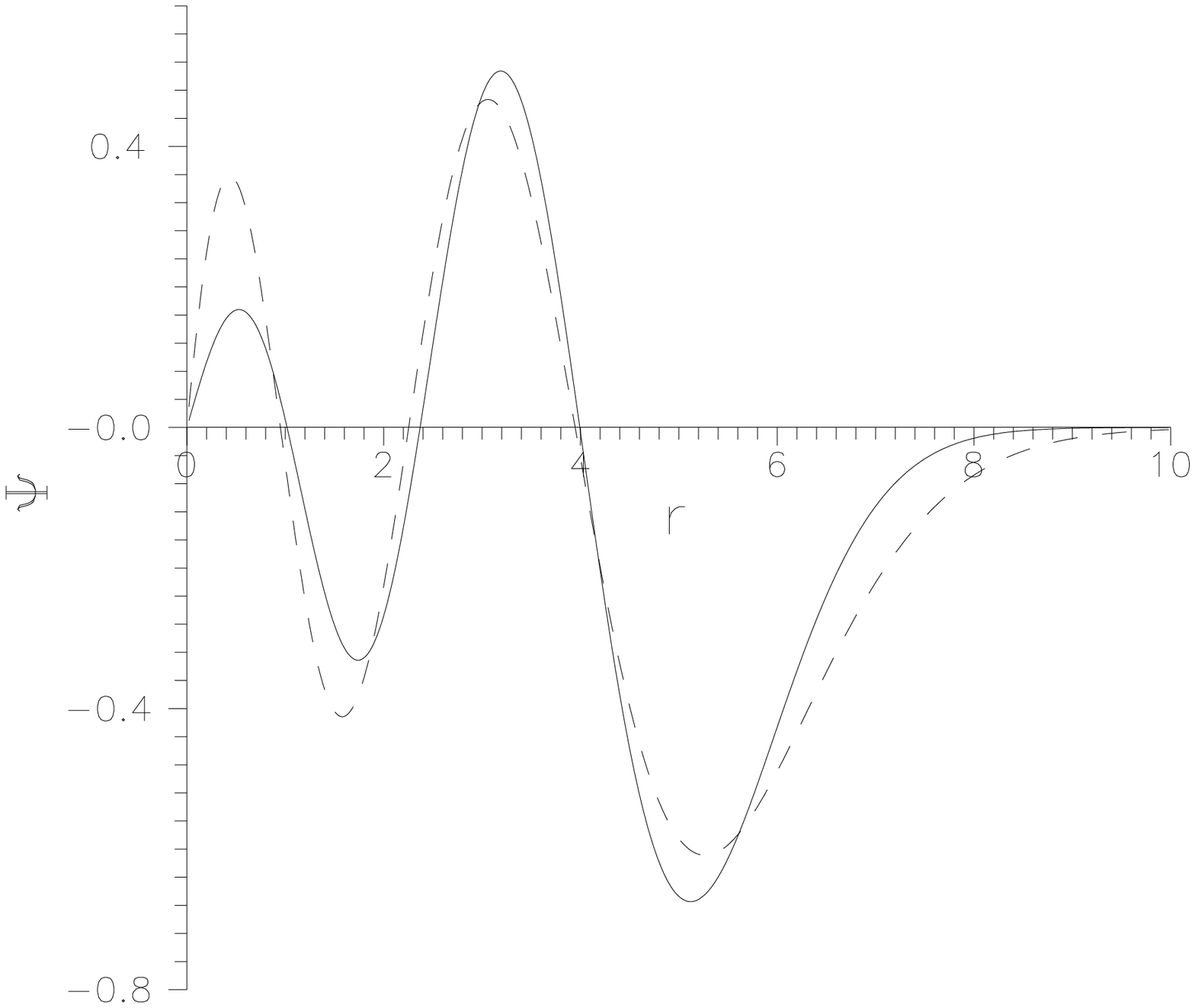}

{ Fig.3. The same as in Fig.2, but for  $\Upsilon (4S)$ and
$k_{\max}=4$. }
\end{center}

\bigskip

 \begin{center}

\includegraphics[width=7cm,keepaspectratio=true]{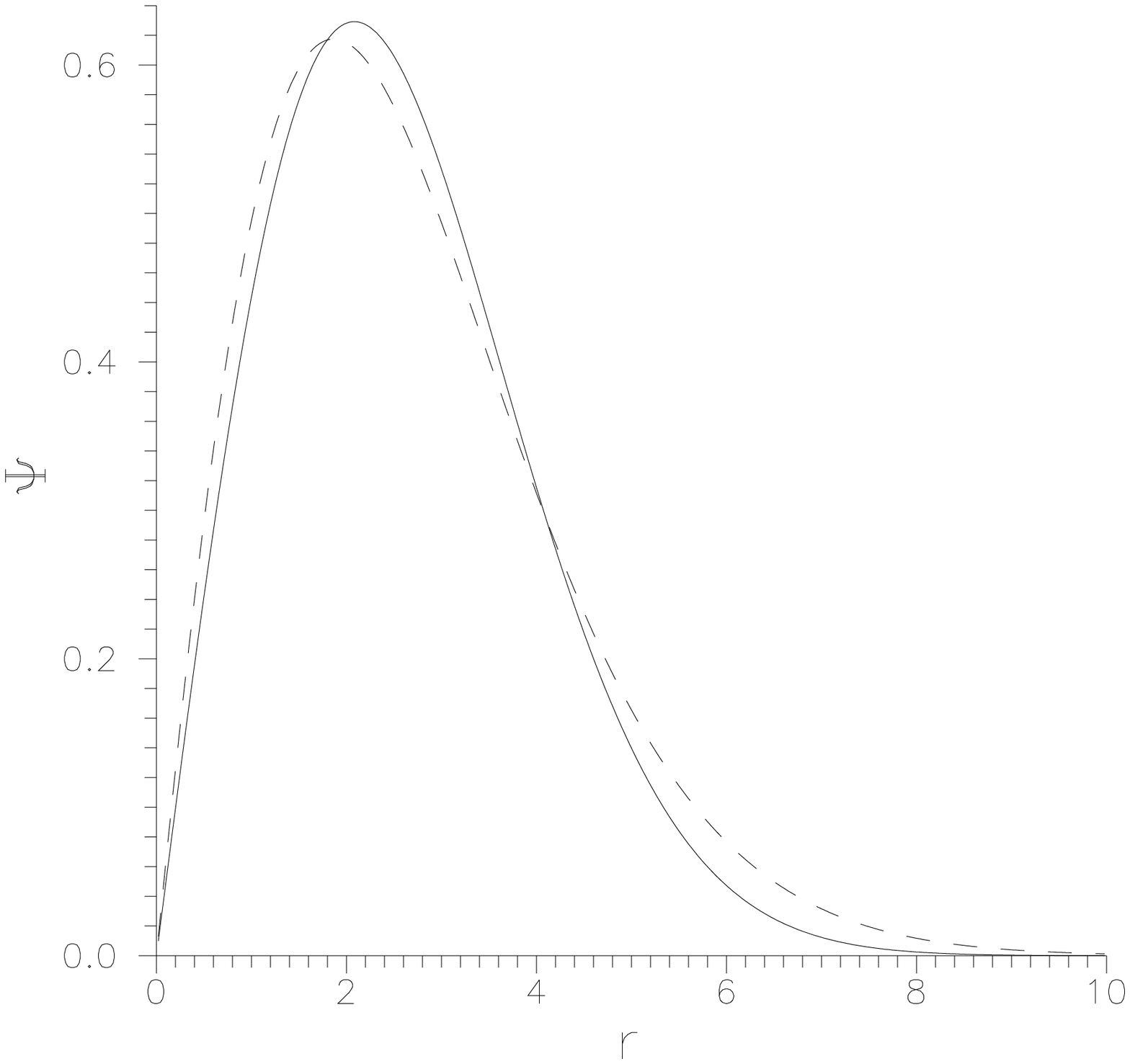}

{ Fig.4. The same as in Fig.2, but for  $B$  meson with
$k_{\max}=1$. }
\end{center}

 \begin{center}

\includegraphics[width=7cm,keepaspectratio=true]{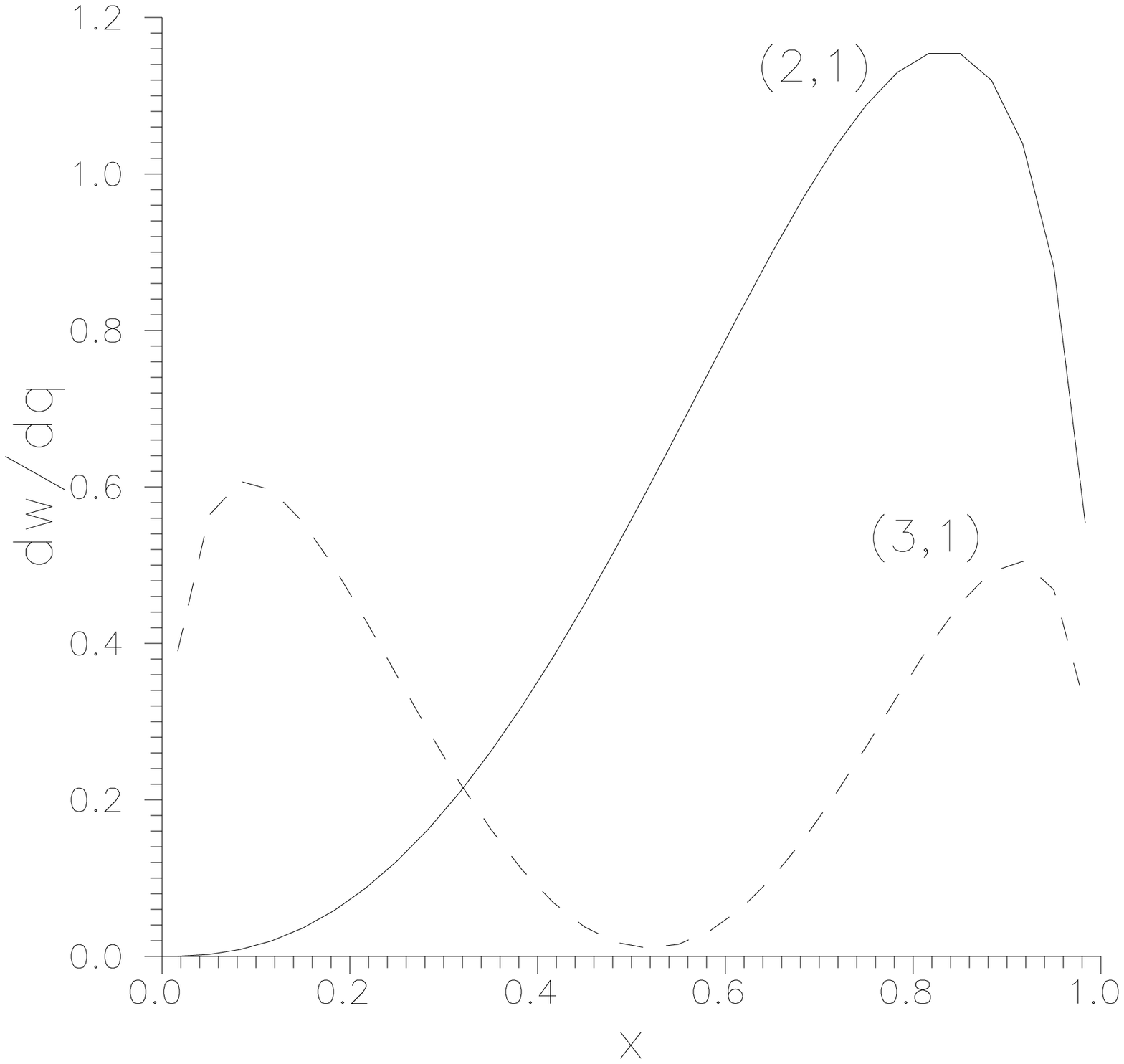}
\end{center}\centerline{ Fig.5. Dipion spectrum
$\frac{dw(21)}{dq}$ as function of $x$ for (2,1) and (3,1)
transitions.}

 \begin{center}

\includegraphics[width=7cm,keepaspectratio=true]{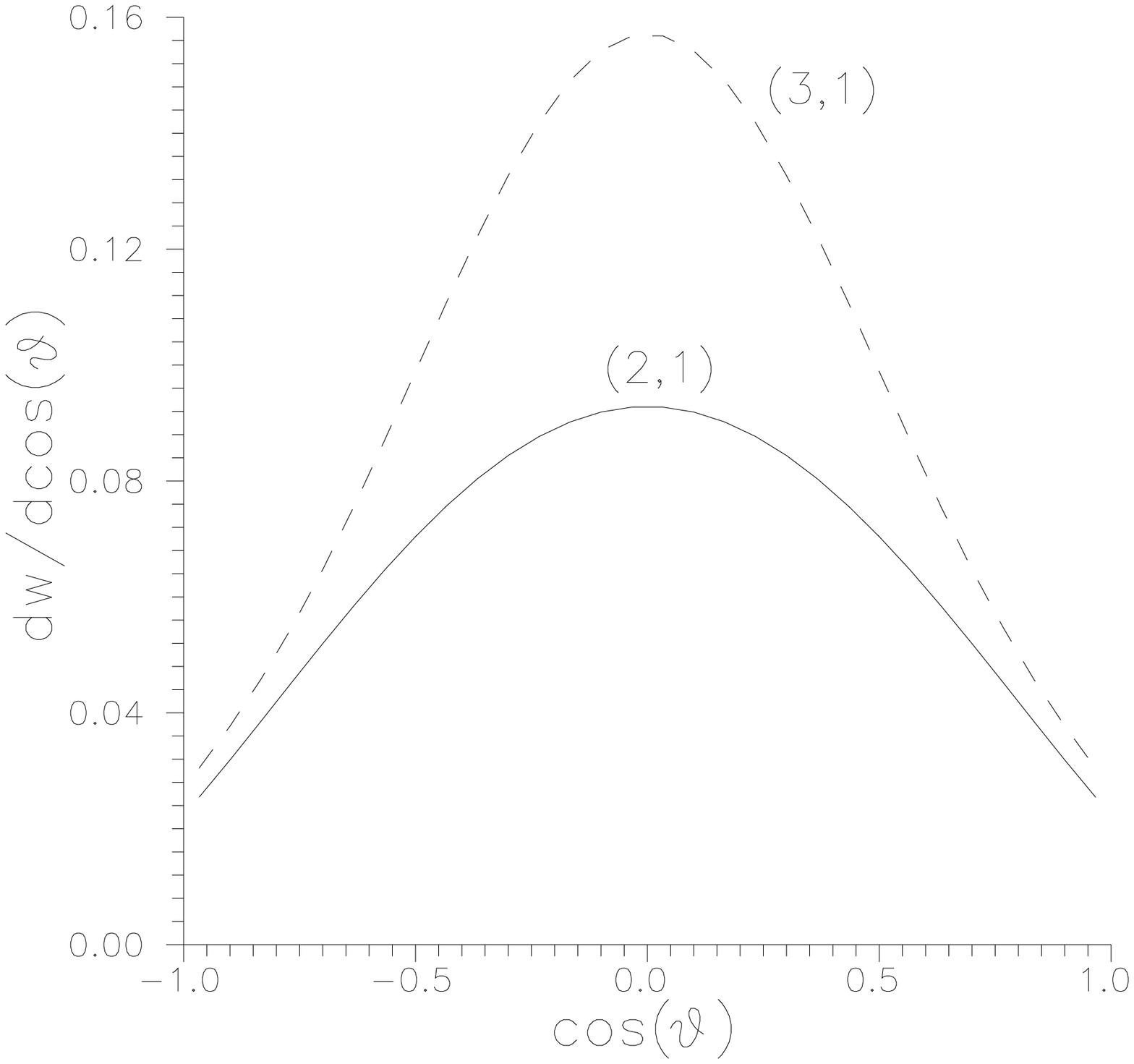}
\end{center}
 \centerline{ Fig.6. Dipion spectrum $\frac{dw(21)}{d\cos\theta}$ as function  of $\cos \theta$ for the (2,1)  and  (3,1)
 transitions.}

\bigskip

\begin{center}

\hspace*{-3cm} \vspace{-1cm}

\includegraphics[width=7cm,keepaspectratio=true]{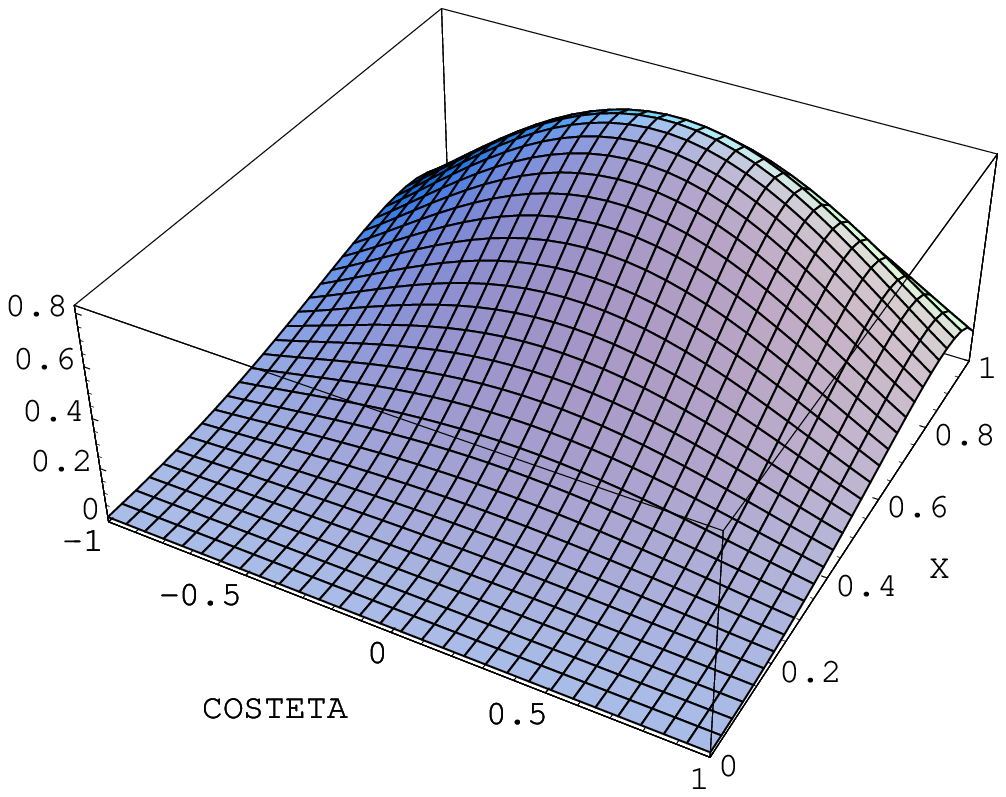}

\end{center}
\centerline{ Fig.7.  The $2d$ dipion spectrum
$\frac{d^2w(2,1)}{dqd\cos \theta}$ as function of $x, \cos \theta$
for the (21) transition.}

\bigskip
 \begin{center}

\includegraphics[width=7cm,keepaspectratio=true]{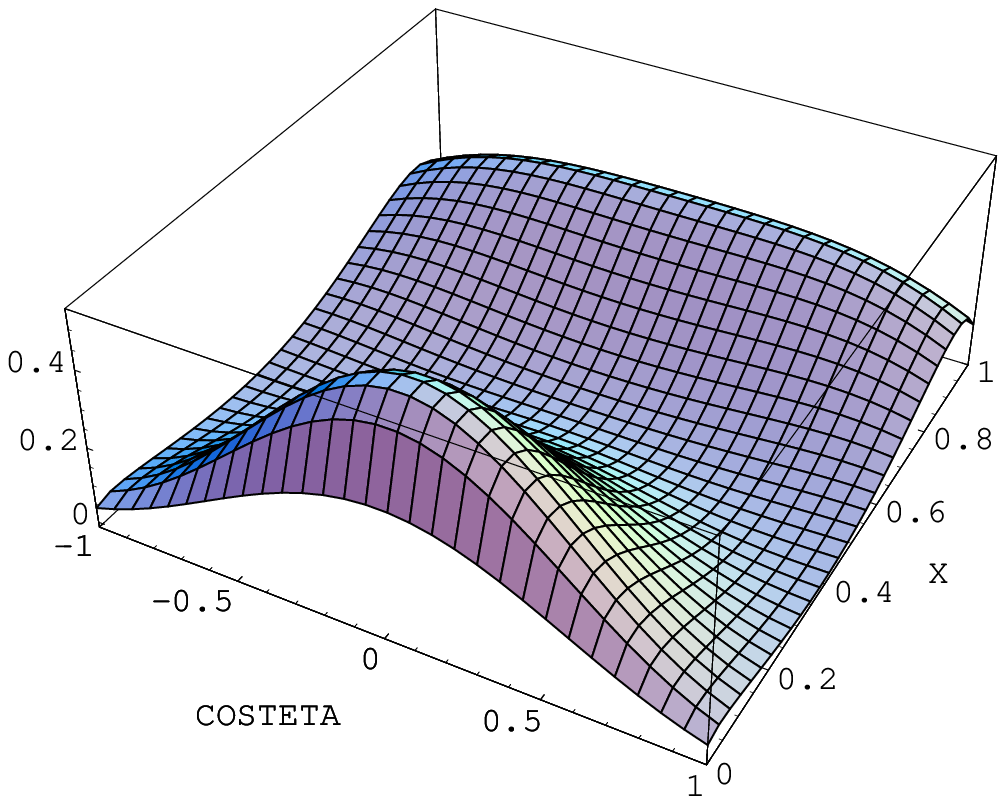}

\end{center}
\centerline{ Fig.8. The same as in Fig.7,  but for the (3,1)
transition. }

 \begin{center}

\includegraphics[width=7cm,keepaspectratio=true]{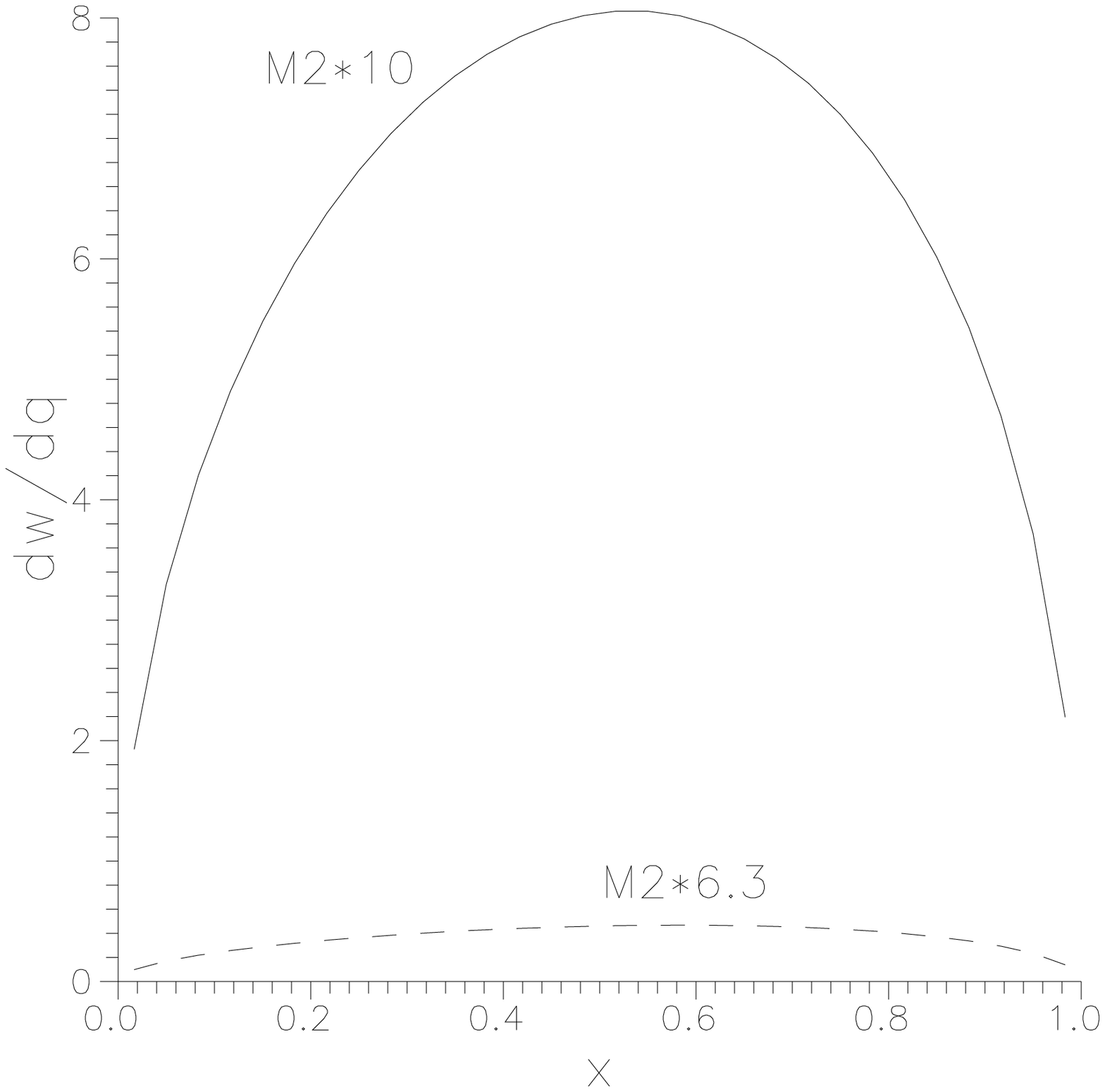}

\end{center}

{ Fig.9. The same as in Fig.5,  but for the (3,2) transition.
Numbers on the curves (6,3) and (10)  refer to the value of ratio
$\frac{M_{\omega}}{M_{br}}$ used in Eq.(18). }

 \begin{center}

\includegraphics[width=7cm,keepaspectratio=true]{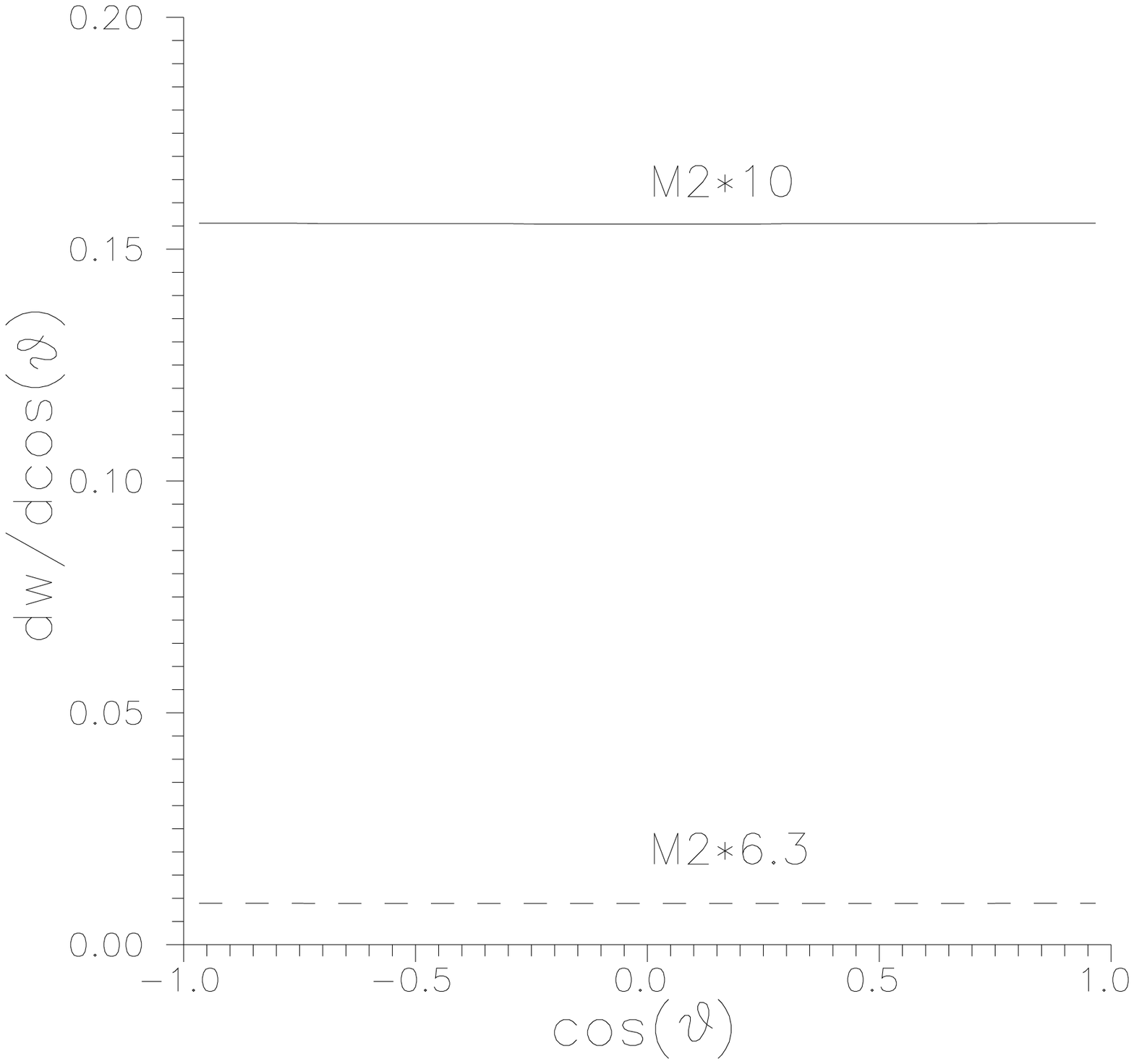}

\end{center}\centerline{  Fig.10. The same as in Fig.9,  but for $\frac{dw}{d\cos \theta}$the (3,2) transition.}

\begin{center}

\hspace*{-3cm} \vspace{-1cm}

\includegraphics[width=7cm,keepaspectratio=true]{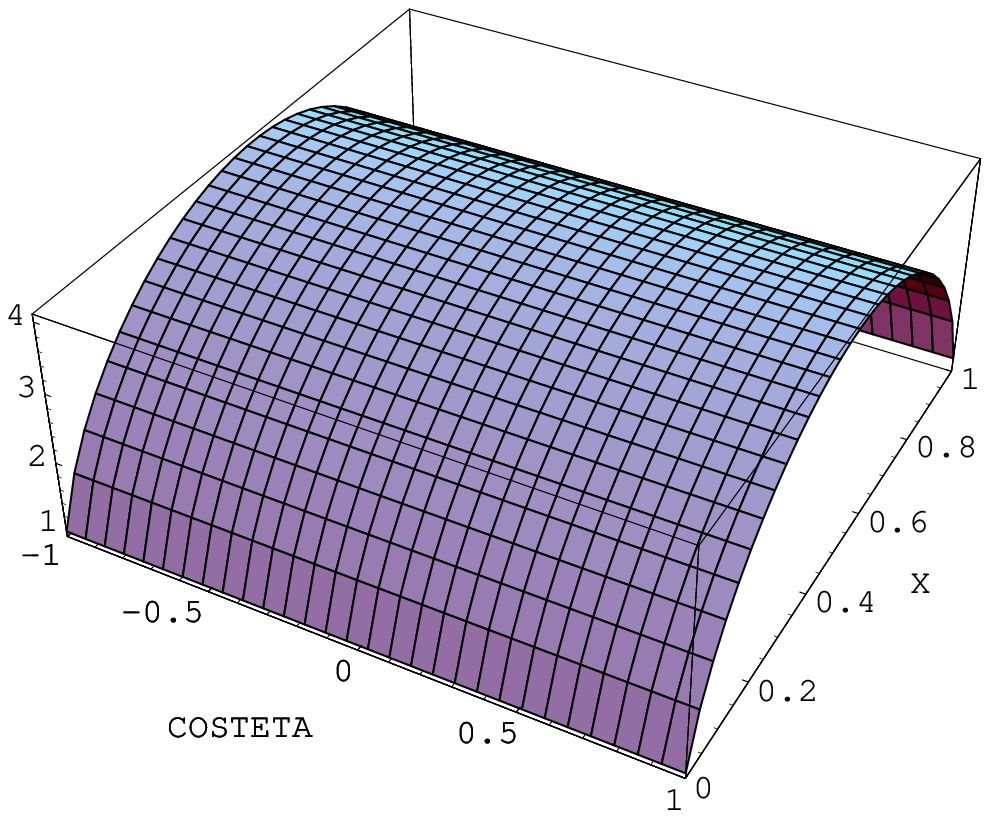}

\end{center}

\centerline{ Fig.11.  The same as in Fig.7,  but for the (3,2)
transition and $\frac{M_{\omega}}{M_{br}}=10$. }

 \begin{center}

\includegraphics[width=7cm,keepaspectratio=true]{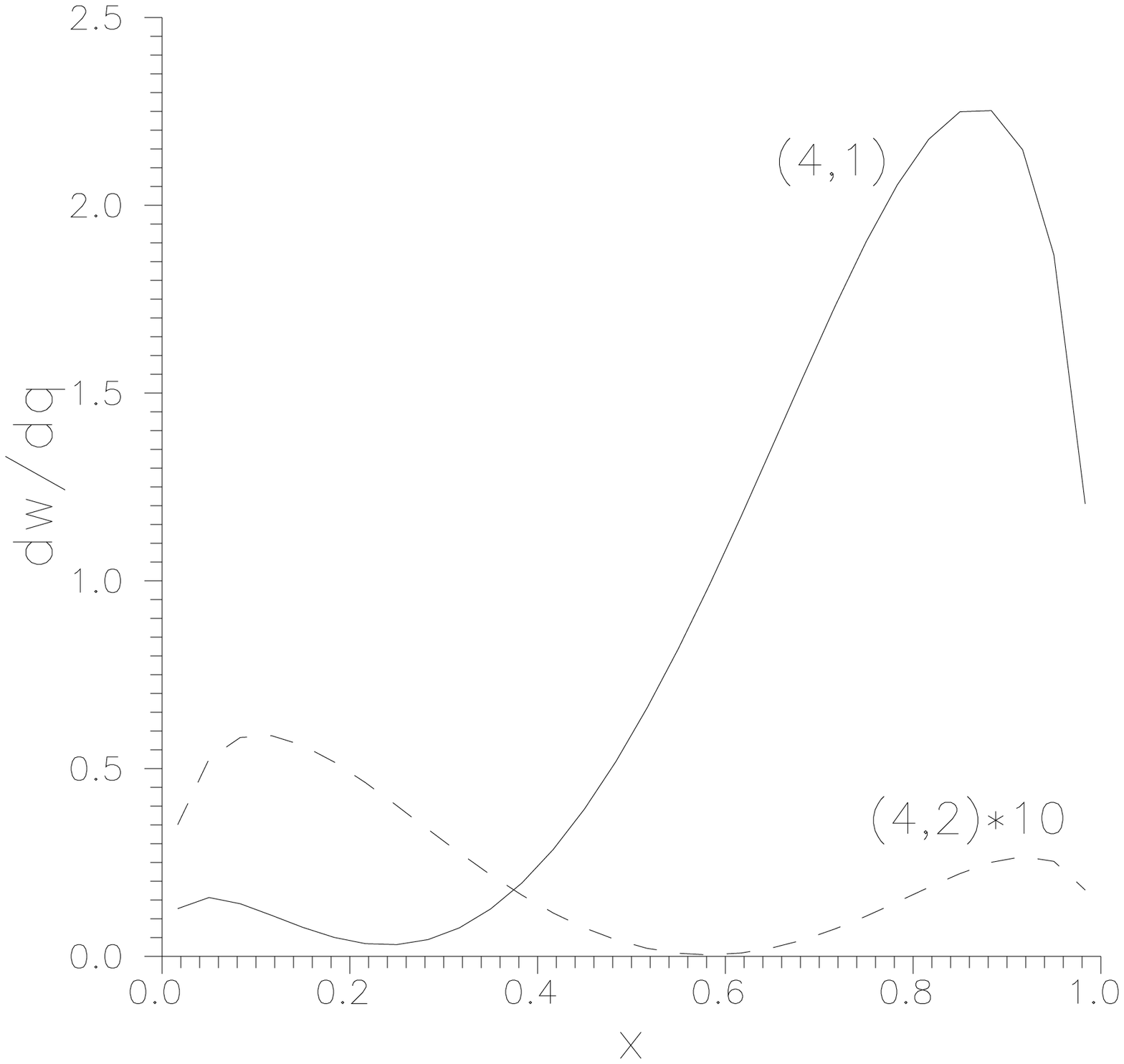}
\end{center}\centerline{  Fig.12. The same as in Fig.5,  but for the (4,1) and (4,2)  transitions.}

 \begin{center}

\includegraphics[width=7cm,keepaspectratio=true]{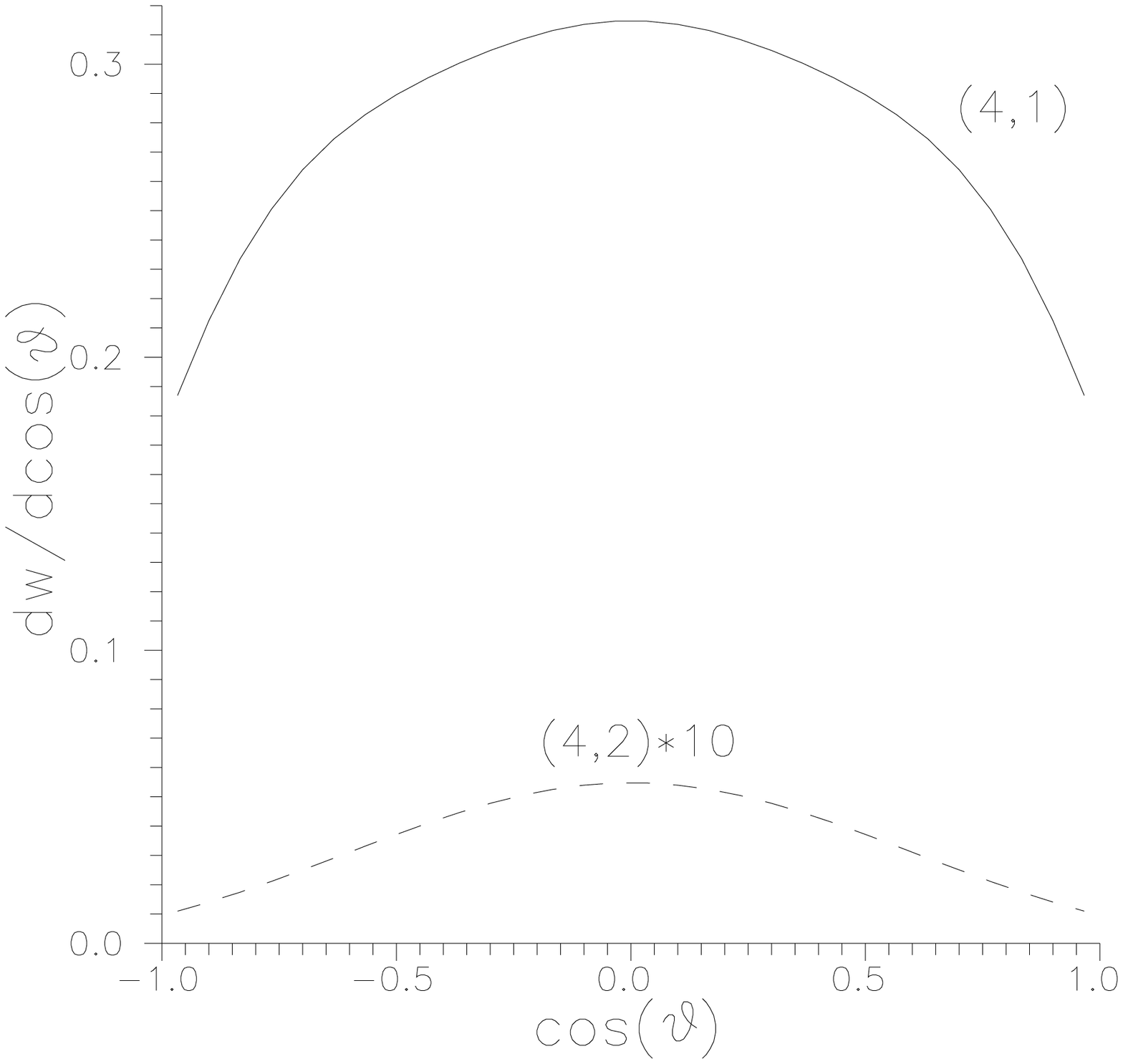}
\end{center} \centerline{ Fig.13.  The same as in Fig.6,  but for the (4,1) and (4,2) transitions.}

\begin{center}

\includegraphics[width=7cm,keepaspectratio=true]{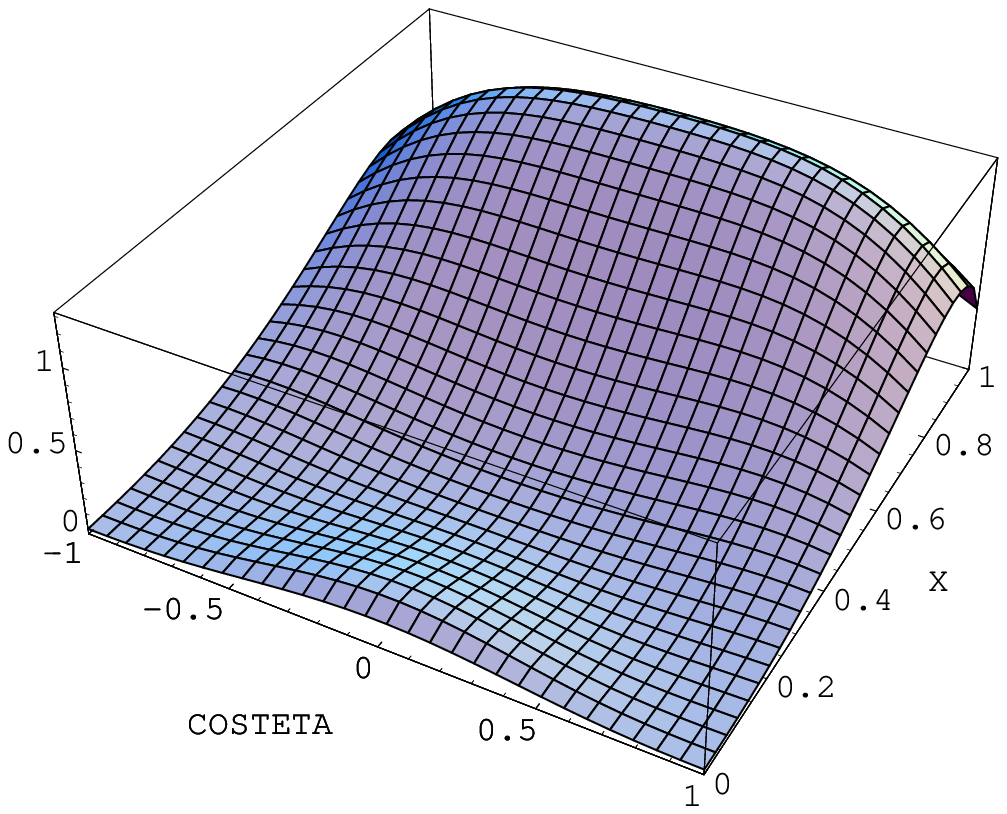}

\end{center}

\centerline{ Fig.14.  The same as in Fig.7,  but for the (4,1)
transition.}

\begin{center}

\includegraphics[width=7cm,keepaspectratio=true]{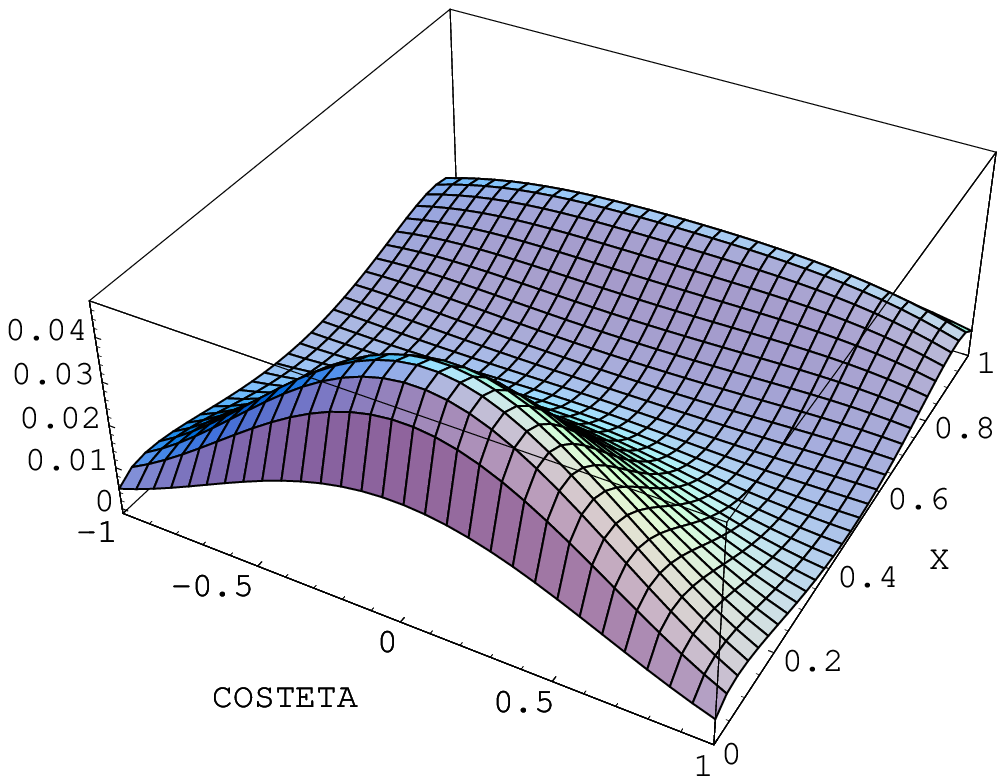}

\end{center}

\centerline{ Fig.15.  The same as in Fig.7,  but for the (4,2)
transition.}


\newpage

 \begin{center}

\includegraphics[width=7cm,keepaspectratio=true]{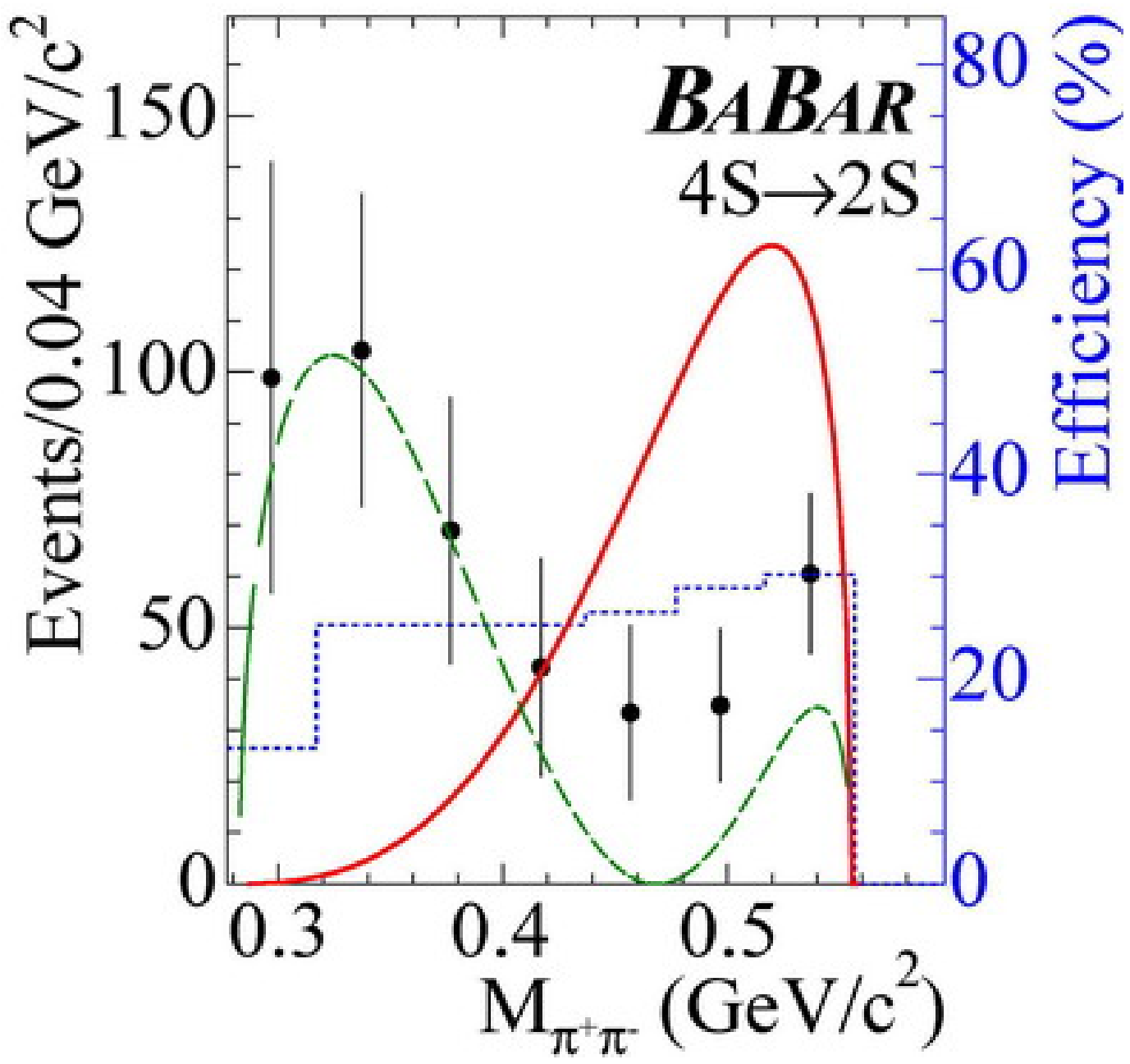}
\centerline{ Fig.16.Comparison of data from [6] with
parametrization (35) (broken line).}

\end{center}

 \begin{center}

\includegraphics[width=8.5cm,keepaspectratio=true]{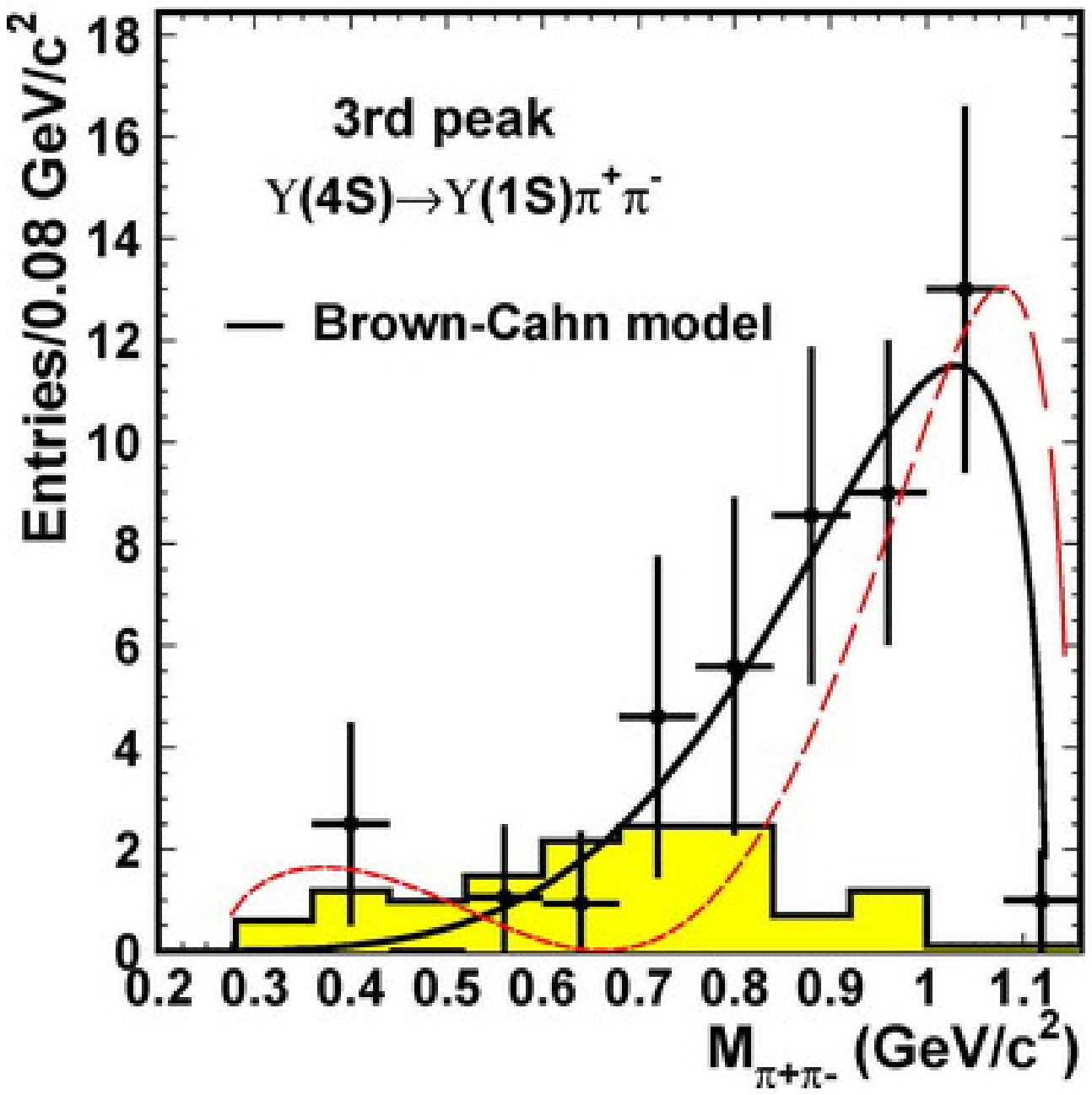}

\centerline{ Fig.17. Comparison of data from [8] with
parametrization (35) (broken line)}
\end{center}

\begin{figure}
\includegraphics[width=16cm,height=20cm,keepaspectratio=true]{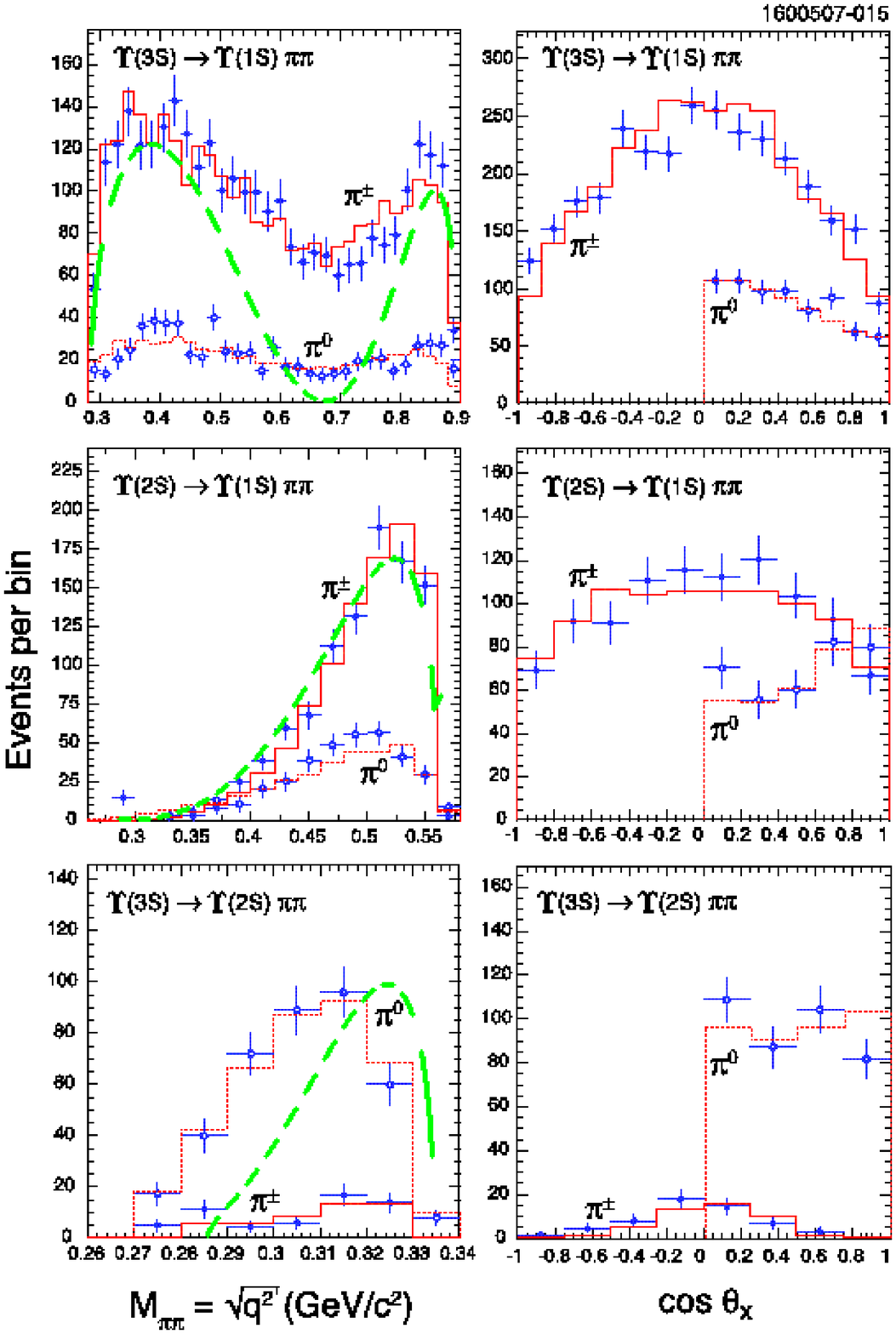}
{  Fig.18. Comparison of data from [6] with parametrization (35)
and parameter $\eta$ from Table 1 (dashed line) (from \cite{31}.}
\end{figure}

 \end{document}